\documentclass[iop]{emulateapj}

\usepackage{epsfig}
\usepackage{epstopdf}
\usepackage{graphicx}
\usepackage{wasysym}
\usepackage{subfigure}
\usepackage{graphics}
\usepackage{longtable}
\usepackage{amssymb}
\usepackage{latexsym} 
\usepackage{wrapfig}
\usepackage{natbib}
\usepackage{wasysym}
\usepackage{mathrsfs}
\usepackage{morefloats}
\shorttitle{Brown Dwarfs from Kernel Phases}
\shortauthors{Pope, Martinache, and Tuthill}

\long\def\symbolfootnote[#1]#2{\begingroup%
\def\thefootnote{\fnsymbol{footnote}}\footnote[#1]{#2}\endgroup}

\begin{document}

\title{Dancing in the Dark: New Brown Dwarf Binaries from Kernel Phase Interferometry\footnotemark[1]}
\author{Benjamin Pope}
\affil{Sydney Institute for Astronomy, School of Physics, University of Sydney, NSW 2226, Australia}
\email{bjsp@physics.usyd.edu.au}
\author{Frantz Martinache}
\affil{National Astronomical Observatory of Japan, Subaru Telescope, Hilo, HI 96720, USA}
\email{frantz@naoj.org}
\author{Peter Tuthill}
\affil{Sydney Institute for Astronomy, School of Physics, University of Sydney, NSW 2226, Australia}
\email{p.tuthill@physics.usyd.edu.au}

\footnotetext[1]{Based on observations performed with the NASA/ESA Hubble Space Telescope. The Hubble observations are associated with proposal ID 10143 and 10879 and were obtained at the Space Telescope Science Institute, which is operated by the Association of Universities for Research in Astronomy, Inc., under NASA contract NAS 5-26555.} 

\begin{abstract}
This paper revisits a sample of ultracool dwarfs in the Solar neighborhood 
previously observed with the Hubble Space Telescope's NICMOS NIC1 instrument.
We have applied a novel high angular resolution data analysis technique 
based on the extraction and fitting of kernel phases to archival data.
This was found to deliver a dramatic improvement over earlier 
analysis methods, permitting a search for companions down to projected
separations of $\sim$1 AU on NIC1 snapshot images. 
We reveal five new close binary candidates and present revised astrometry 
on previously-known binaries, all of which were recovered with the technique. 
The new candidate binaries have sufficiently close separation to determine dynamical 
masses in a short-term observing campaign. 
We also present four marginal detections of objects  
which may be very close binaries or high contrast companions.
Including only confident detections within 19 parsecs,
we report a binary fraction of at least $\epsilon_b = 17.2^{+5.7}_{-3.7} \%$.
The results reported here provide new insights into the population of nearby 
ultracool binaries, while also offering an incisive case study of the benefits 
conferred by the kernel phase approach in the recovery of companions within a
few resolution elements of the PSF core.
\end{abstract}

\keywords{techniques: interferometric --- techniques: image processing  --- techniques: high angular resolution --- 
  stars: low-mass --- stars: formation --- (stars:) brown dwarfs}

\section{Introduction}
\label{intro}

A detailed picture of multiplicity is an essential element to understanding low-mass 
stars and brown dwarfs. Binary systems present an opportunity to determine 
model-independent dynamical masses when both astrometry and radial velocity data
are available. Systems so characterized may then become part of the foundations for 
the construction of an observationally constrained mass-luminosity-age sequence 
applicable across the entire class.

Furthermore, the statistical properties of populations of low mass binaries have
profound implications on the basic physics of star formation and solar system assembly. 
Multiplicity rates are a key discriminant between hypotheses about the formation and 
evolution of low mass systems, as discussed in \citet{2007prpl.conf..427B}. 
Two main mechanisms have been proposed for the formation of brown dwarfs 
in the field: embryo ejection, and gravoturbulent collapse \citep{Basu06072012}. 
Specifically, the embryo ejection hypothesis predicts a low binarity incidence 
($\sim 8\%$) \citep{2012MNRAS.419.3115B}, which conflicts with the observed binarity rate 
($\sim 15\%$) \citep{2005nlds.book.....R}. 
Mapping the incidence of binarity, and in particular extending completeness
to smaller orbital separation is therefore of interest in establishing the primary 
formation mechanism of field brown dwarfs.

Snapshot imaging is a straightforward way to discover new multiple
systems. Intrinsically faint and red, L-dwarfs present challenging targets
for ground based observations, typically requiring laser guide star (LGS)
adaptive optics (AO). 
Space telescopes naturally offer high Strehl ratio imaging at
diffraction-limited resolution, with the major departures from ideal performance
arising from field-dependent PSF changes, spacecraft jitter, and slow optical
drift from thermally induced breathing modes of the mechanical structures.

Imaging campaigns with the HST have demonstrated notable success in prospecting
for companions to cool objects, providing high quality diffraction-limited images 
of a large number of targets \citep{2006AJ....132..891R,2008AJ....135..580R}. 
These campaigns have shed light on the population of cool dwarfs in the 
Solar neighborhood.

The simplest and most widely used method for detection of companions in snapshot 
imaging essentially relies on direct visual examination of images. 
Obvious companions are quickly identified, and traditional astronomical image 
analysis tools, namely aperture photometry and centroiding, provide the important 
astrometric and photometric characteristics of the target. 
Faint or close-in companions are, however, easily missed in a visual search and 
identifying such objects requires more sophisticated computational techniques. 

For example, some stellar images exhibit an elongation along one axis as noted 
by \citet{2006AJ....132..891R} which may be suggestive of the presence of a barely-resolved 
companion.
Subtraction of a model PSF has been exploited to infer the presence of 
a companion \citep{1998PASP..110.1046K,2004ApJ...617.1323P,2012AJ....144...64D}, although the performance of this approach is
arguably poor, and furthermore it weakly constrains the relative 
photometry and astrometry. We propose to look at the same images from an interferometric
standpoint, leveraging the exquisite level of calibration this
technique offers.

For the detection and characterization of companions at
small angular separations, non-redundant masking (NRM) interferometry
used in conjunction with AO has demonstrated outstanding performance, e.g. in  
\citet{2006SPIE.6272E.103T,2006ApJ...650L.131L,2012ApJ...745....5K}.
The key underpinning such successes has been the robust, self-calibrating nature of the
observables recovered from NRM interferometry, and in particular the {\it closure phase},
first suggested for the radio \citep{1958MNRAS.118..276J} and later exploited in the optical 
\citep{1986Natur.320..595B}.
Imaging systems where the phase on any given baseline in the pupil is disturbed by 
random phase errors from atmospheric or instrumental aberrations suffer from degraded
performance. 
However by summing phases around closed loops of non-redundant baselines, these random phasors 
cancel out and the resulting closure phases are extremely robust to wavefront aberrations.
NRM interferometry from the ground \citep{2000PASP..112..555T} relies heavily on closure phase 
for high-contrast detection, and there are plans to extend the technique to space platforms
\citep{2009SPIE.7440E..30S}.
Recent observations achieved with this technique reported by
\citet{2006ApJ...650L.131L, 2007ApJ...661..496M, 2008ApJ...678..463I,
  2008ApJ...679..762K} and  \citet{2009ApJ...695.1183M}
demonstrate that the level of calibration achieved with interferometric measurements 
permits the detection of companions at scales at or even somewhat beyond the diffraction 
limit of the imaging system. 
Recently, NRM interferometry succeeded in providing evidence for a low-luminosity companion
in the transitional disk host systems T~Cha \citep{2011A&A...528L...7H} and LkCa~15 \citep{2012ApJ...745....5K}.

It has recently been demonstrated that if a conventional (full-aperture) PSF is of sufficient 
quality (wavefront residual errors typically $\lesssim\lambda/4$), an analogous set of high-quality 
interferometric observables can be extracted from the images \citep{2010ApJ...724..464M}.
These new quantities are the {\it kernel phases}, and represent a generalization of the idea of 
closure phase to a redundant pupil configuration.
The major advance offered by kernel phase interferometry is that it is not restricted to 
non-redundant pupils. 
In brief, for small wavefront errors (i.e. high Strehl ratio), the phase errors in the pupil 
plane can be related to those in the Fourier plane by a linear operator. 
The kernel or null-space of this operator therefore singles out a subspace of baseline phases 
which are not affected by this error, which can then play the same role as closure phases 
in providing a robust set of observables to constrain image structure.
Kernel phases were first successfully extracted from HST/NICMOS data
on a single target by \citet{2010ApJ...724..464M}, demonstrating
significant improvement over more traditional data analysis
\citep{2004ApJ...617.1323P}. The technique has now also been
successfully applied to ground based AO observations
\citep{2011SPIE.8151E..33M}.

This paper revisits a sample of nearby ultracool dwarfs observed by
the Hubble Space Telescope NICMOS NIC1 camera and first presented in
\citet{2006AJ....132..891R} and \citet{2008AJ....135..580R}.
Our analysis allows dramatic extensions to the discovery space for putative
companions, and in particular explores separation ranges down to 1 AU on 
targets located within 20\,parsec.
Section~\ref{sec:methods} provides an overview of the dataset and
introduces the methods used for our new analysis. 
Section~\ref{sec:res} discusses the results of the kernel phase analysis for
the entire sample and implications for the astrophysical interpretation of
brown dwarf formation.

\section{Sample and Methods}
\label{sec:methods}

\subsection{Sample of Ultracool Dwarfs}
\label{sec:sample}

This study focuses on two samples of ultracool dwarfs, observed with the
HST/NICMOS NIC1 camera, and whose properties were reported by
\citet{2006AJ....132..891R} and \citet{2008AJ....135..580R}. Each target was observed in two filters: F110W and F170M, which correspond loosely to the astronomical $J$ and $H$ bands. These differ in that the $J$ and $H$ bands sample atmospheric transmission windows, which do not constrain space-based observations. We will use $J$ and $H$ as shorthands for F110W and F170M respectively, but the difference should be noted. Table~\ref{tbl:sample} summarizes the observational properties of the combined sample as stated in Table 1 of \citet{2006AJ....132..891R} and Table 1 of \citet{2008AJ....135..580R}.

In addition to detecting several binaries by traditional data analysis methods, these authors also provide a list of 43 and 26 apparently unresolved objects in the 2006 and 2008 samples respectively, which we revisit in this paper.
All ten of the previously resolved binaries were independently recovered with kernel phases, and for all we report significantly improved astrometric precision. In addition to confirming the technique and software on an unambiguous sample, the dramatic improvements to the binary parameters offer the chance to determine orbital elements and therefore dynamical masses. Both the detections and the remaining unresolved binaries are used in quantitative exploration of the performance limitations of kernel phase analysis in the recovery of high contrast systems.

\subsection{Kernel Phase Analysis}
\label{sec:kpanalysis}

\begin{figure}
\plotone{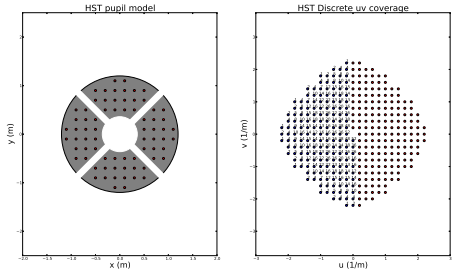}
\caption{Diagrams showing (left) the discretized pupil model used for kernel phase analysis and (right) the resultant $(u,v)$ sampling points.}
\label{fig:model}
\end{figure}

Kernel phase analysis follows the principles introduced in
\citet{2010ApJ...724..464M}. 
The first step is to generate a model of the pupil of the imaging system as 
seen from the detector. 
This task is straightforward based on information contained in the TinyTim v.~7.2 PSF
simulation package for NIC1 \citep{2011SPIE.8127E..16K}, available at \texttt{tinytim.stsci.edu}. 
For kernel phase analysis, the model pupil is discretized into a square grid 
array of 72 sub-apertures with a unit spacing $1/12$\,th that of the pupil 
diameter (cf. Figure~\ref{fig:model}).
Regions of the primary blocked by spiders or the secondary mirror are not
sampled, and one can also observe that the unit baseline imposed by this sampling 
of the pupil imposes an outer working angle of 6\,$\lambda/D$.

This geometry fills the $(u,v)$-plane with a regular grid of 176 distinct sample 
points at a cadence of 12 points across the diameter.
The transfer matrix that relates instrumental phase errors to spurious $(u,v)$-phase 
information is therefore a $72 \times 176$ rectangular matrix, whose SVD reveals 36  
non-zero singular values (that is exactly one half of the entire number of sample 
points in the pupil), leaving $176 - 36 = 140$ independent kernel phase relations. 
Kernel phase analysis can therefore recover $140/176 = 80\%$ of the available phase 
information present in the quantized grid. 

The specific discretization chosen was found to have relatively little impact on
the performance of the algorithm. 
If instead we adopt a finer-sampled pupil model with 20 points across the diameter, 
we get 1516 kernel phases out of 1632 distinct baselines, leading to a $93\%$ phase 
recovery. 
We analyzed a portion of our dataset with this finer 20 point sampling and found 
little improvement in the quality of fit or precision in parameter estimation, though 
we note somewhat better agreement between $H$ and $J$ bands with the finer pupil model using 
Levenberg-Marquardt model fitting. 
Because the finer grid analysis was computationally expensive but yielded only a small 
change in fitted binary parameters, it was judged that application over a large grid to 
fit the available data was unwarranted. 
We have therefore chosen the coarser model for our fitting routines, but note that 
more computer time may produce some improvements with a finer pupil model. 
For application to wider separation binaries, however, the finer model would be 
strictly required: if we have $p$ points across the pupil, Nyquist's sampling theorem 
imposes an outer working angle ${p \lambda}/{2 D}$. 
If this condition is not met, the Fourier plane fringes will not be well-sampled and 
parameter estimates will be subject to aliasing or may not be recovered at all.

A shorter wavelength of observation delivers an increase of angular resolution, but 
with the same level of optical aberration, this also precipitates a greater degree of 
image degradation (lower Strehl ratio).  
When considering residual phase noise, we therefore expect that the kernel phase 
signal-to-noise will be accordingly higher for images taken at longer wavelengths. 

We are furthermore limited by the fact that we only have single snapshots of each 
target: without multiple frames it is difficult to calibrate systematic errors and 
explore statistical uncertainties on the kernel phase observables. 
We therefore selected a sample of stars for which we could see no PSF abnormality or 
obvious Fourier phase structure, and repeatedly applied a Levenberg-Marquardt fitting 
algorithm to the raw kernel phase data to attempt to find binary companions. 
Those targets for which no companion model was significantly preferred over a single
source were deemed to be ``unresolved''. 
We then used this unresolved population to establish uncertainties as ensemble standard 
deviations for each kernel phase, which in turn enables quantification of significance 
in subsequent explorations entailing $\chi^2$ fitting. 
The results presented here could be considerably improved with the design of an
observational campaign at the outset which delivers better diversity, by exploiting multiple 
exposures and dedicated point source calibrators.
A more comprehensive understanding of systematic errors and noise estimates for individual 
targets would yield more sensitive limits on detection and better errors on fitted 
parameters. 

\subsection{Bayesian Methods}
\label{bayes}

A binary system at any one epoch can be characterized by its angular separation $\delta$, position angle $\theta$ and contrast ratio $c$. The likelihood of a binary model with these parameters given the set of kernel phases $\{K\phi_j\}$ is related to the $\chi^2$ statistic by

\begin{equation}
  L(\delta,\theta,c|\{K\phi_j\}) \propto \exp (-\chi^2/2).
  \label{eq:likely}
\end{equation}

When normalized, this likelihood is the joint density probability
function for all three parameters. 
When calculating $\chi^2$, we found it necessary to add an additional systematic error term in 
quadrature to bring the minimum reduced $\chi^2$ down to 1. Confidence intervals for any
individual parameter can be calculated by integrating over the two
other parameters. After this {\it marginalization}, we estimate the parameter and its
uncertainty from the mean and the standard deviation of the
1-D marginal distribution respectively.

The approach closely follows established practice with
closure-phase in NRM-interferometry for the characterization of
binaries \citep{2009ApJ...695.1183M}. 
When applied to our sample HST imaging data set, the final results from our algorithm 
were: (1), confirmation of binaries already identified with other methods; 
(2), the determination of statistically sound constraints on the binary parameters; and 
(3), a robust statistical estimate for the probability that signals extracted from any given
system betray the presence of a companion or can be attributed to noise.

The sampling and integration of the likelihood function given in Equation~\ref{eq:likely} is in general difficult, and is typically performed by a computationally-expensive grid integration or a Monte Carlo Markov Chain (MCMC) random sampling method. In this paper we apply a recently developed alternative, namely nested sampling. This method, proposed by \citet{2004AIPC..735..395S}, uses an unusual change of variables to calculate the model evidence. It has recently seen a surge of interest; e.g. for cosmological model fitting \citep{2006ApJ...638L..51M,2008IJMPA..23..787M}, and the analysis of simulated gravitational wave data \citep{2009CQGra..26k4011A,2009CQGra..26u5003F}.

The key idea of nested sampling is to populate the allowed prior space with a large number ($\sim 100$) of `active points' 
which are initially chosen at random and subsequently evolved towards ensemble states of successively higher likelihood using
MCMC methods. 
Our implementation was based on \citet{sivia2006data} and ultimately yielded a statistical representation of the likelihood space 
which could be used for binary hypothesis testing and estimation of model parameter values and their uncertainties.
Although a number of alternate gradient-descent and MCMC methods were benchmarked, nested sampling was found to be 
computationally the most efficient.
A global binarity analysis of the entire sample in both the $J$ and $H$ bands could be accomplished quickly, however
for objects which are in the barely resolved limit, there are well known strong parameter degeneracies -- particularly between 
separation and brightness of a companion. 
This ambiguity conflates bright close companion models with somewhat more distant fainter companion models, considerably diminishing
the astrophysical utility of the findings.

In addition to separately fitting image data in $J$ and $H$ bands, nested sampling was fast enough to enable joint 
four-parameter fitting of both images simultaneously $\mathcal{M}(\delta,\theta,c_H,c_J)$. 
The ambiguity in separation/contrast from separate fitting was found to be greatly ameliorated by 
the covariance of separation with contrast. 
Joint fitting enforces identical separation between bands, greatly restricting the size of the $\chi^2$ valley of 
degeneracy with contrast ratio. 
These findings are promising for the coming generation of Integral Field Unit cameras working with extreme AO 
systems which naturally deliver spatio-spectral data cube observations.

For the joint fitting, an additional error term was added in quadrature to represent unknown noise sources. 
This was found iteratively such that each band separately had a minimum reduced $\chi^2$ of 1 at the best joint fit 
parameters.
In cases where the existing error estimates resulted in a minimum reduced $\chi^2 < 1$, no adjustment was made. 

For this study, we searched a delimited parameter space for companions.
We initially searched up to a contrast ratio of 200, somewhat beyond the limits established in Section \ref{deteclimits}. 
Candidate binaries were then compared to Plots~\ref{cojointsig} and \ref{cojointsignear} to establish significance.
The range of separation explored ran from 30 to 200\,mas and all position angles were considered.

\small

\begin{deluxetable}{lcccc}
\tablecolumns{5}
\tablewidth{0pc}
\tablecaption{Sample of unresolved L-dwarfs from \citet{2006AJ....132..891R} and \citet{2008AJ....135..580R} (after line break)}
\tablehead{
  \colhead{2 MASS name} & \colhead{Sp. type} & 
  \colhead{J} & \colhead{H} & \colhead{K}
}
\startdata
2MASS J00361617+1821104  &L3.5     &12.47    &11.59    &11.06\\
2MASS J00452143+1634446  &L0       &13.06    &12.06    &11.37\\
2MASS J01075242+0041563  &L8       &15.82    &14.51    &13.71\\
2MASS J01235905-4240073  &M8       &13.15    &12.47    &12.04\\
2MASS J01550354+0950003  &L5       &14.82    &13.76    &13.14\\
2MASS J02132880+4444453  &L1.5     &13.51    &12.77    &12.24\\
2MASS J03140344+1603056  &L0       &12.53    &11.82    &11.24\\
2MASS J03552337+1133437  &L6       &14.05    &12.53    &11.53\\
2MASS J04390101-2353083  &L6.5     &14.41    &13.37    &12.81\\
2MASS J04455387-3048204  &L2       &13.41    &12.57    &11.98\\
2MASS J05002100+0330501  &L4       &13.67    &12.68    &12.06\\
2MASS J05233822-1403022  &L2.5     &13.12    &12.22    &11.63\\
2MASS J06244595-4521548  &L5       &14.48    &13.34    &12.60\\
2MASS J06523073+4710348  &L4.5     &13.55    &12.37    &11.69\\
2MASS J08251968+2115521  &L7.5     &15.12    &13.79    &13.05\\
2MASS J08354256-0819237  &L5       &13.15    &11.95    &11.16\\
2MASS J08472872-1532372  &L2       &13.52    &12.63    &12.05\\
2MASS J09083803+5032088  &L7       &14.56    &13.47    &12.92\\
2MASS J09111297+7401081  &L0       &12.92    &12.20    &11.75\\
2MASS J09211410-2104446  &L2       &12.78    &12.15    &11.69\\
2MASS J10452400-0149576  &L1       &13.13    &12.37    &11.81\\
2MASS J10484281+0111580  &L1       &12.92    &12.14    &11.62\\
2MASS J10511900+5613086  &L2       &13.24    &12.42    &11.90\\
2MASS J11040127+1959217  &L4       &14.46    &13.48    &12.98\\
2MASS J11083081+6830169  &L0.5     &13.14    &12.23    &11.60\\
2MASS J12130336-0432437  &L5       &14.67    &13.68    &13.00\\
2MASS J12212770+0257198  &L0       &13.17    &12.41    &11.95\\
2MASS J14283132+5923354  &L5       &14.78    &13.88    &13.27\\
2MASS J14482563+1031590  &L5       &14.56    &13.43    &12.68\\
2MASS J15074769-1627386  &L5       &12.82    &11.90    &11.30\\
2MASS J15394189-05200428 &L3.5     &13.92    &13.06    &12.58\\
2MASS J15525906+2948485  &L1       &13.48    &12.61    &12.03\\
2MASS J16580380+7027015  &L1       &13.31    &12.54    &11.92\\
2MASS J17054834-0516462  &L0.5     &13.31    &12.54    &12.03\\
2MASS J17312974+2721233  &L0       &12.09    &11.39    &10.91\\
2MASS J17534518-6559559  &L4       &14.10    &13.11    &12.42\\
2MASS J18071593+5015316  &L1.5     &12.96    &12.15    &11.61\\
2MASS J19360262-5502367  &L4       &14.49    &13.63    &13.05\\
2MASS J20575409-0252302  &L1.5     &13.12    &12.27    &11.75\\
2MASS J21041491-1037369  &L2.5     &13.84    &12.96    &12.36\\
2MASS J22244381-0158521  &L4.5     &14.05    &12.80    &12.01\\
2MASS J23254530+4251488  &L7.0     &15.51    &14.46    &13.81\\
2MASS J23515044-2537367  &L0.5     &12.46    &11.73    &11.29\\
& & & & \\
2MASS J002424.6-015819	&M9.5	&11.86	&11.12	&10.58\\
2MASS J010921.7+294925	&M9.5	&12.91	&12.16	&11.68\\
2MASS J022842.4+163933	&L0	&13.17	&12.33	&11.82\\
2MASS J025114.8-035245	&L3	&13.08	&12.26	&11.65\\
2MASS J025503.5-470050	&L8	&13.23	&12.19	&11.53\\
2MASS J031854.0-342129	&L7	&15.53	&14.31	&13.48\\
2MASS J044337.6+000205	&M9	&12.52	&11.80	&11.17\\
2MASS J083008.3+482848	&L8	&15.44	&14.34	&13.68\\
2MASS J085925.4-194926	&L7	&15.51	&14.44	&13.73\\
2MASS J102248.2+582545	&L1	&13.50	&12.64	&12.16\\
2MASS J102552.3+321235	&L7	&15.91:	&15.59:	&15.07\\
2MASS J104307.5+222523	&L8	&15.95	&14.75	&13.99\\
2MASS J105847.8-154817	&L3	&14.18	&13.24	&12.51\\
2MASS J115539.5-372735	&L2	&12.81	&12.04	&11.46\\
2MASS J120358.1+001550	&L4	&14.01	&13.06	&12.48\\
2MASS J130042.5+191235	&L1	&12.71	&12.07	&11.61\\
2MASS J142131.5+182741	&L0	&13.23	&12.43	&11.94\\
2MASS J142528.0-365023	&L3	&13.75	&12.58	&14.49\\
2MASS J143928.4+192915	&L1	&12.76	&12.04	&11.55\\
2MASS J150654.4+132106	&L3	&13.41	&12.41	&11.75\\
2MASS J151500.9+484739	&L6	&14.06	&13.07	&12.57\\
2MASS J172103.9+334415	&L3	&13.58	&12.92	&12.47\\
2MASS J200250.7-052152	&L6	&15.32	&14.23	&13.36\\
2MASS J202820.4+005227	&L3	&14.30	&12.38	&12.79\\
2MASS J214816.3+400359	&L6.5	&14.15	&12.78	&11.77\\
2MASS J223732.5+392239	&M9.5	&13.35	&12.68	&12.15
\enddata
\label{tbl:sample}
\end{deluxetable}
\normalsize

\section{Results and Discussion}
\label{sec:res}

Images for all objects in Table~\ref{tbl:sample} in the two filter bands (F110W and F170M)
were recovered in digital form from the HST MAST Archive, where they are listed under Proposals 10143 and 10879. 
All data were processed using our kernel phase techniques, and in the discussion which
follows, we divide our results into three subsections: \ref{known}, binaries already reported; 
\ref{discovery}, new binaries uncovered by kernel phase; \ref{highcon} marginal detections meriting further study; and \ref{deteclimits},  sample detection thresholds and the incidence of unresolved sources.

\subsection{Known binaries}
\label{known}

In all cases where companions were reported by \citet{2006AJ....132..891R} 
or \citet{2008AJ....135..580R}, our new analysis independently 
recovered strong systematic signals confirming binarity. 
We therefore confirm all previously reported detections, and we stress that
our analysis was blind in the sense that no prior knowledge was employed
in our search.

In common with closure phases, non-zero excursions in the kernel phases encode
information about asymmetric structures, although the abstracted nature of
the observable makes for significant challenges in intuitive data presentation.
One approach to present the way a binary signal is encoded upon the
kernel phases, and the fingerprint match of this complex function to the 
actual recorded data, is to simply plot the best-fit model quantities against 
the observed data in a correlation diagram.

\begin{figure*}[h]
\plottwo{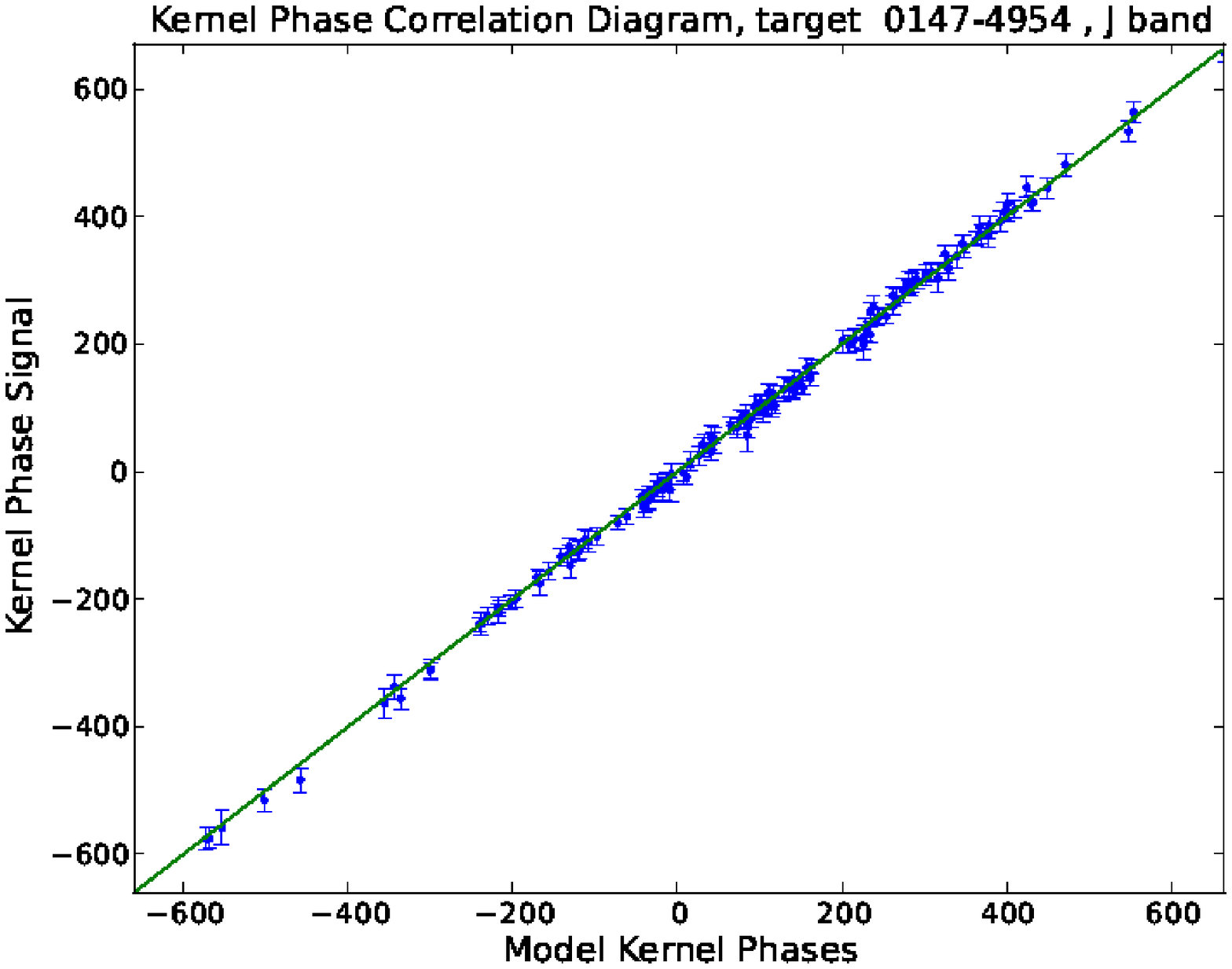}{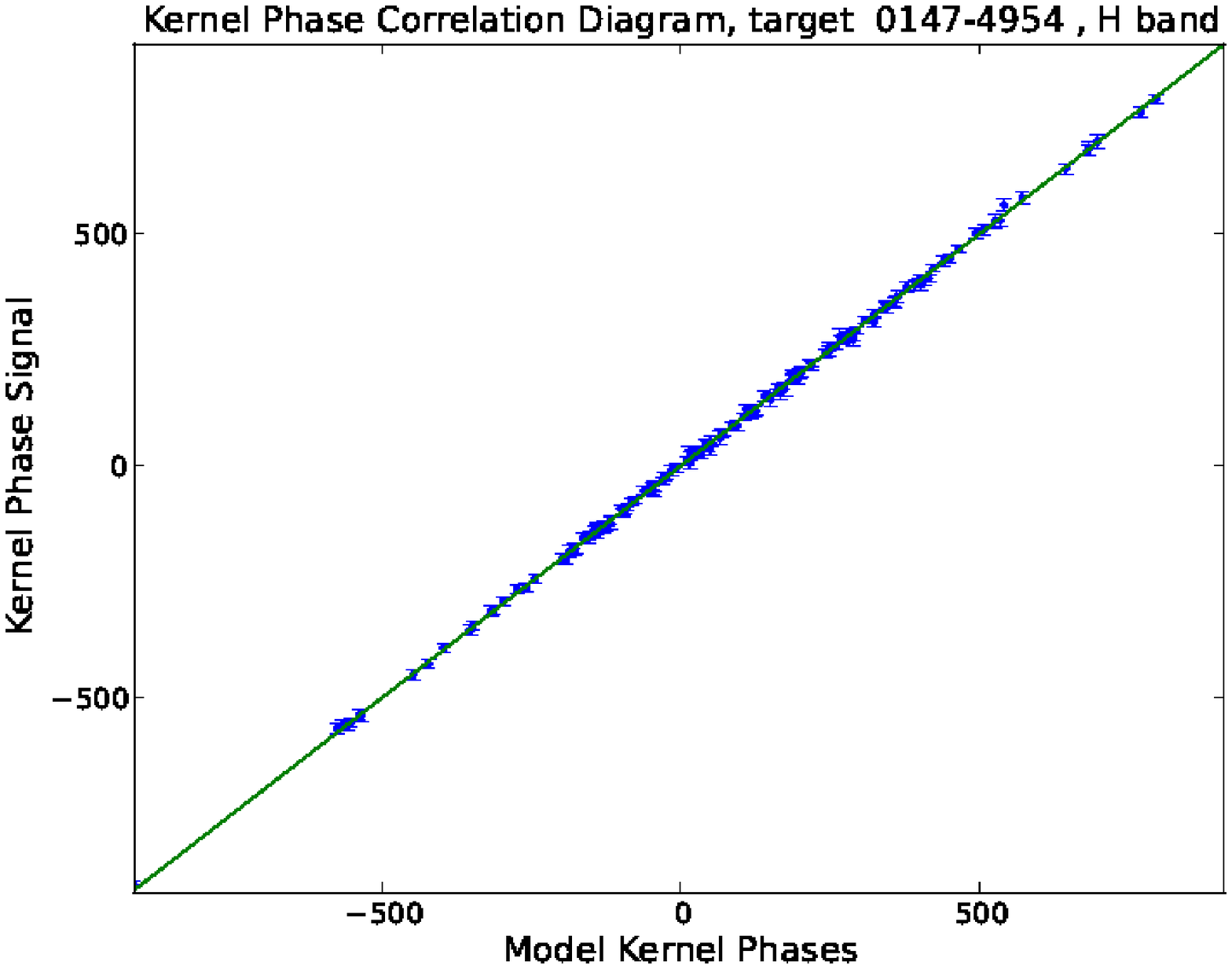}
\caption{Correlation diagrams for the previously-detected binary 2M 0147-4954 in (left) $J$ band and (right) $H$ band.}
\label{59corr}
\end{figure*}

\begin{figure*}[h]
\plottwo{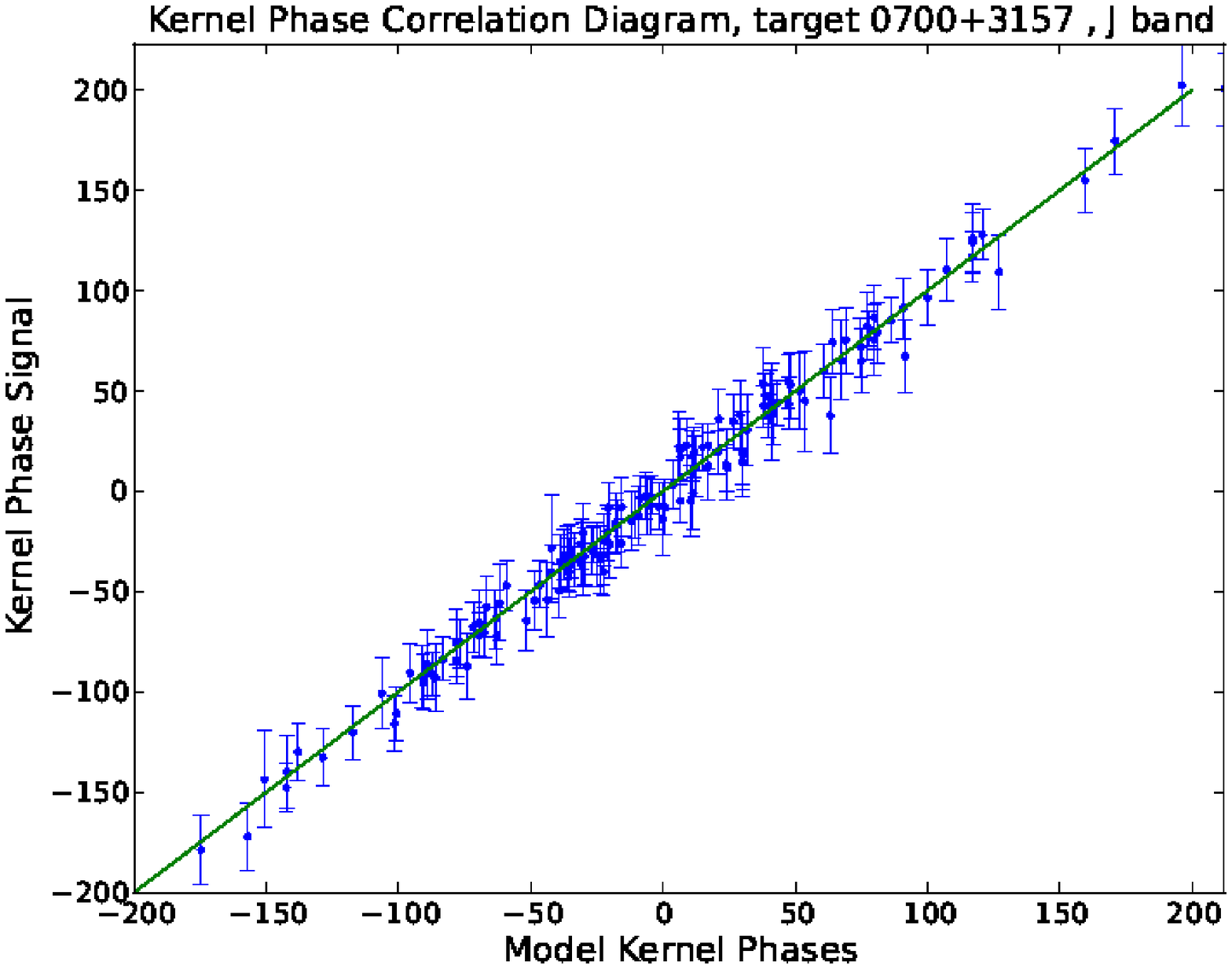}{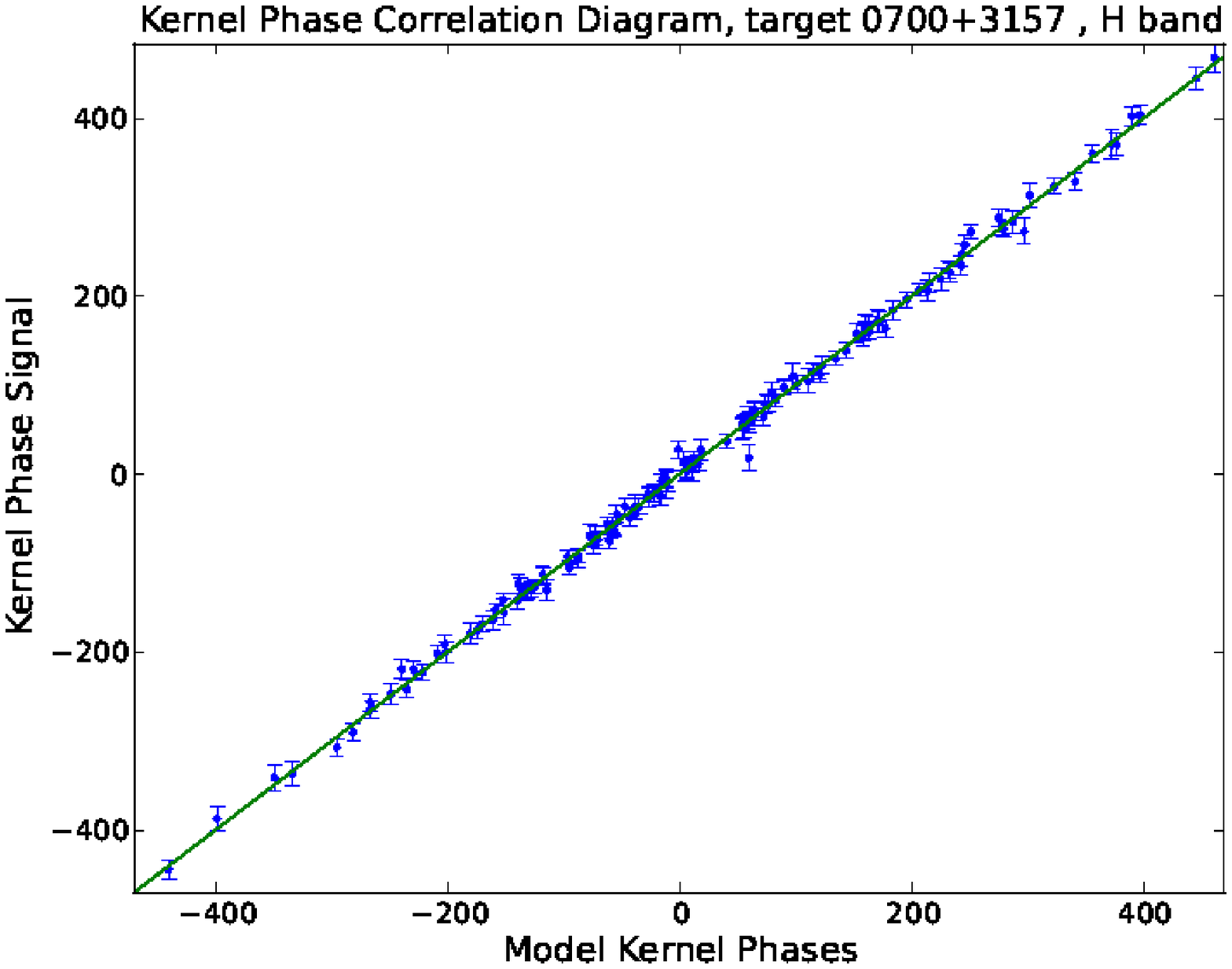}
\caption{Correlation diagrams for the previously-detected binary 2M 0700+3157 in (left) $J$ band and (right) $H$ band.}
\label{19corr}
\end{figure*}

Figures~\ref{19corr} and \ref{59corr} present such a diagram of the binary 
model fit against extracted kernel phase data for two illustrative cases: 
previously-known binaries 2M~0700+3157 and 2M~0147-4954.
The one-to-one correspondence line is overplotted, delineating the locus of 
perfect fit.

Kernel phase analysis has yielded greatly improved astrometric precision
on most previously known binary systems. 
Whereas previous studies, relying on visual analysis and PSF subtraction, 
quoted separations to the nearest 10 mas and position angle to $\sim 1^\circ$, 
kernel phase delivers about one order of magnitude better precision.
All fitted binary parameters agree, typically to within $1\sigma$, when separate
fits to $J$ and $H$ band kernel phases are computed. 
Table~\ref{oldbins} gives final best fit parameters from the simultaneous $J$/$H$ 
four-parameter fit. 
In many cases, the best fits differ from previously published estimates, sometimes 
very substantially (although formal errors were not quoted in 
\citet{2006AJ....132..891R} or \citet{2008AJ....135..580R}).

Counterintuitively, for some of the most readily apparent binaries 
(e.g. 2M~0004-4044, 0025+4759, 0915-0422 and 2152+0937), kernel phase methods 
proved problematic and could fail to converge to a good fit. 
This was particularly true for well-separated, low contrast systems which are
most easily discerned by simple inspection.
Wide binaries induce phase curvature not well sampled by our pupil model,
and for these cases, a direct fit to the squared visibilities was performed.
Visibility data were calibrated by dividing by the ensemble means over the sample, 
with dimensionless errors of $\sim 0.05$ added in quadrature.
As with the kernel phases, statistical analysis was based on nested sampling and
the results also yielded overwhelming improvements in astrometric precision.

The PSF of the individual target 2M~0004-4044 was truncated at the edge of the image, 
presumably due to spacecraft mis-pointing, and therefore did not permit a useful 
kernel phase fit in $H$ band. Nevertheless, a $J$ band kernel fit was found to agree well with 
the parameters published in \citet{2006AJ....132..891R}, and likewise a joint visibility fit agreed well in both bands. The results of the visibility fit are quoted in Table~\ref{oldbins}.
It is unclear, however, what effect the data edge truncation may have on our interferometric observables and therefore the values given in Table~\ref{oldbins} are likely to be subject to an additional unknown error for this system. 

\begin{figure}[h]
\plotone{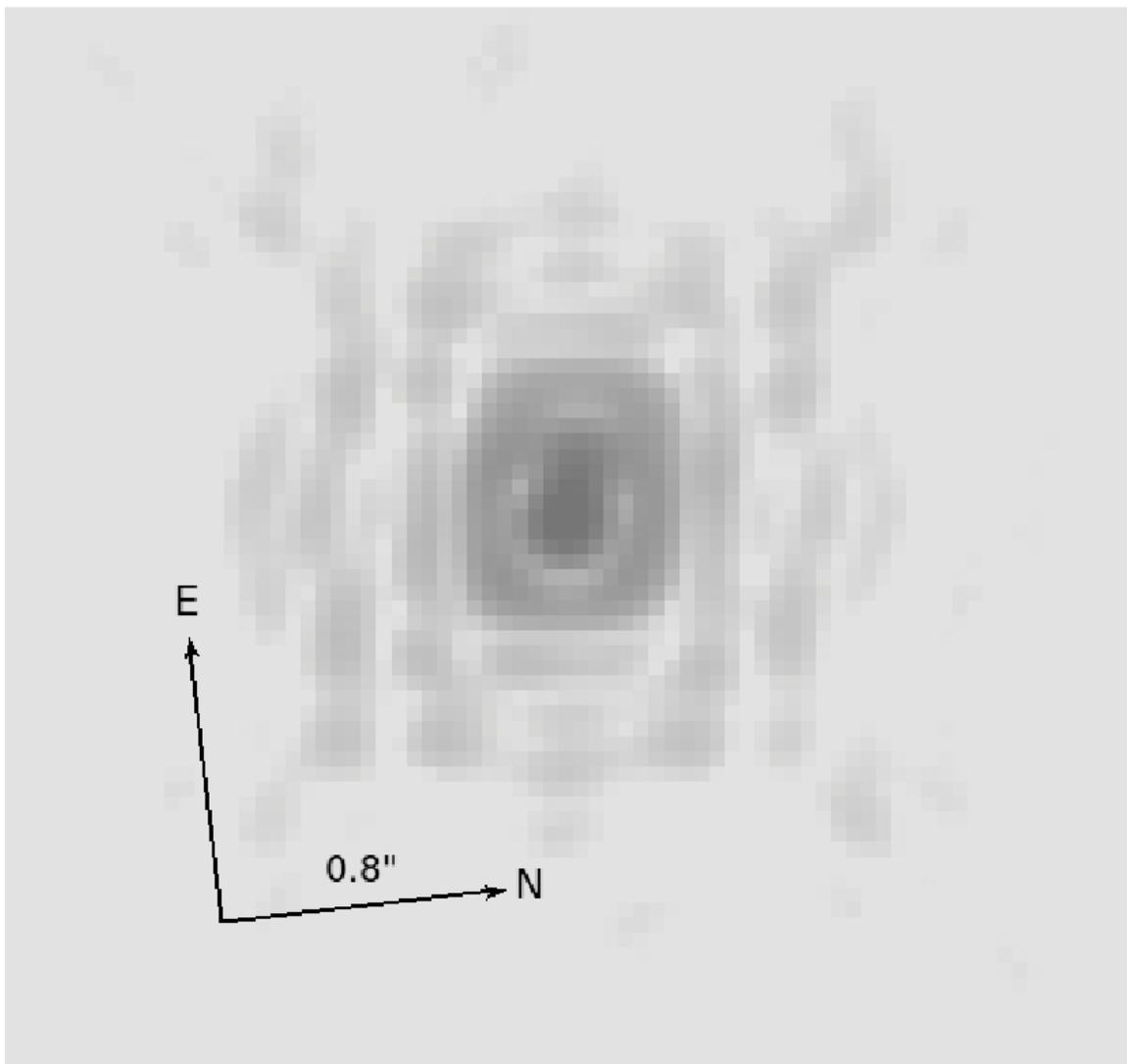}
\caption{Log scale image of 2M 0147-4954 in F170M filter. The corresponding correlation diagrams are displayed in Figure~\ref{59corr}. Note that while the PSFs are hard to distinguish visually, the kernel phase fit is excellent.}
\label{59pic}
\end{figure}

Example correlation diagrams and a corresponding NICMOS image are shown in Figure~\ref{59pic} and Figure~\ref{59corr}. Note that while the PSFs of the primary and companion overlap and are difficult to visually distinguish, they permit a clear kernel phase fit with very precise parameter estimates.

\begin{deluxetable*}{lcccc}
\tablecolumns{5}
\tabletypesize{\small}
\tablecaption{Model Parameters for Known L-Dwarf Binaries}
\tablewidth{0pt}
\tablehead{
  \colhead{2 MASS} &\colhead{Sep.} & 
  \colhead{Pos. angle} & \colhead{Contrast} & \colhead{Contrast} \\
  \colhead{Number} &\colhead{(mas)} & 
  \colhead{(degrees)} & \colhead{Ratio (J)} & \colhead{Ratio (H)}
}
\startdata
0004-4044\tablenotemark{a}  & 84.6$\pm$0.2 & 224.6$\pm$0.1 &1.04$\pm$0.01 & 1.02$\pm$0.02    \\ 
0025+4759\tablenotemark{a}  & 329.0$\pm$0.3 & 233.04$\pm$0.06 & 1.32$\pm$0.04 & 1.03$\pm$0.04 \\ 
0147-4954                   & 139.8$\pm$0.1 & 72.66$\pm$0.05 & 2.37$\pm$0.09 & 2.06$\pm$0.06   \\ 
0429-3123                   & 525.2$\pm$1.2 & 285.3$\pm$0.2 & 3.51$\pm$0.1 & 2.82$\pm$0.06  \\ 
0700+3157                   & 179.6$\pm$0.5 & 105.8$\pm$0.1 & 4.52$\pm$0.07 & 3.81$\pm$0.03  \\ 
0915-0422\tablenotemark{a}  & 738.6$\pm$0.15 & 26.89$\pm$0.01 & 1.114$\pm$0.002  & 1.264$\pm$0.002  \\ 
1707-0558                   & 1009.5$\pm$1.0 & 34.9$\pm$0.05 & 10.6$\pm$0.15 & 7.5$\pm$0.2   \\ 
2152+0937\tablenotemark{a}  & 253.7$\pm$0.09 & 94.5$\pm$0.02 & 1.09$\pm$0.03 & 1.15$\pm$0.03  \\ 
2252-1730\tablenotemark{b}  & 125.9$\pm$0.4 & 353.5$\pm$0.1 & 2.46$\pm$0.01 & 3.395$\pm$0.03  \\ 
2255-5713                   & 178.6$\pm$0.4 & 172.7$\pm$0.1 & 5.05$\pm$0.08 & 4.33$\pm$0.02  \\ 
\enddata
\tablenotetext{a}{~Low contrast: fit with visibilities. See Section \ref{known}.}
\tablenotetext{b}{~This object is the subject of \citet{2006ApJ...639.1114R}.}
\label{oldbins}
\end{deluxetable*}


\subsection{Discovery of New Binary Candidates}
\label{discovery}

\begin{figure*}[h]
\plottwo{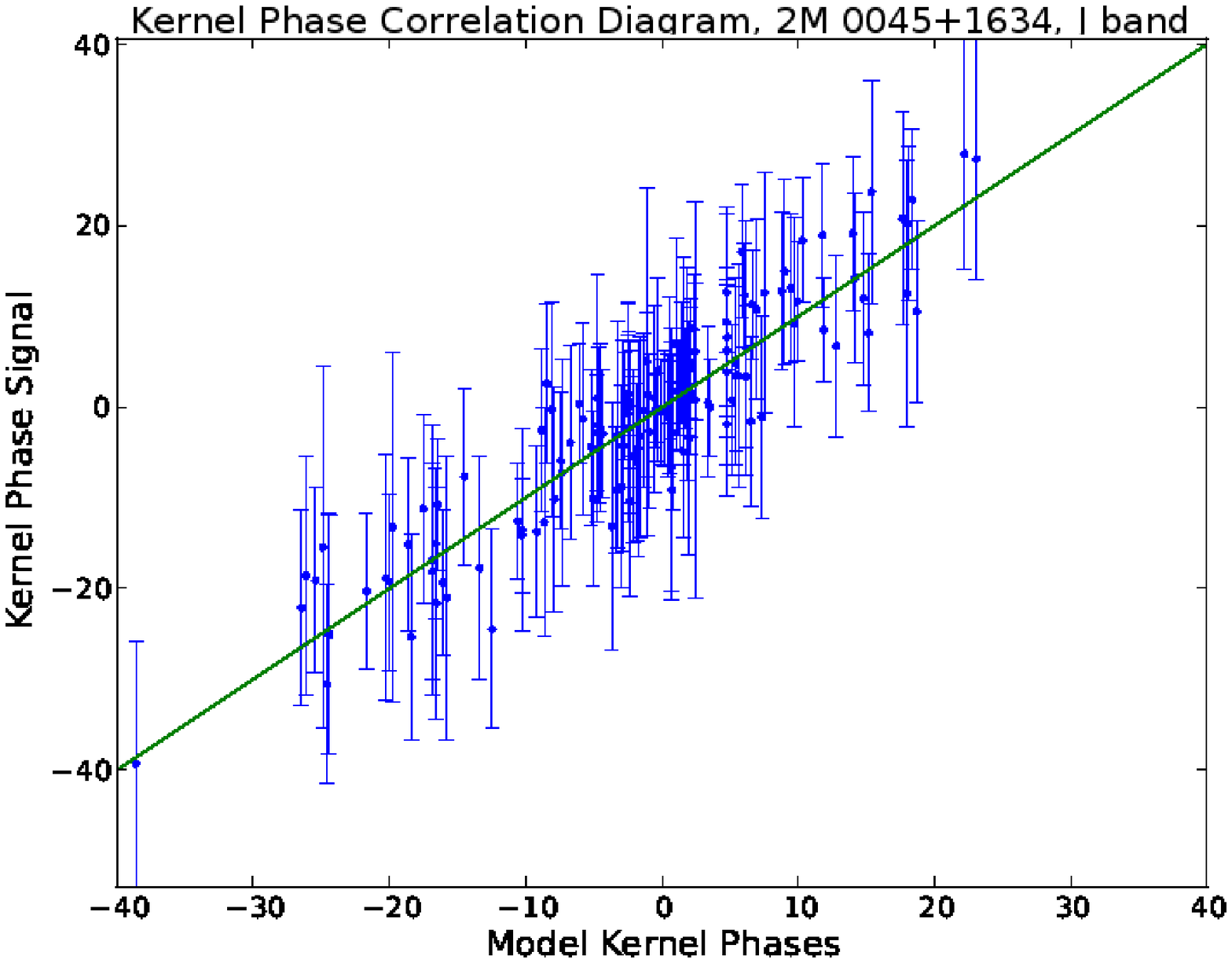}{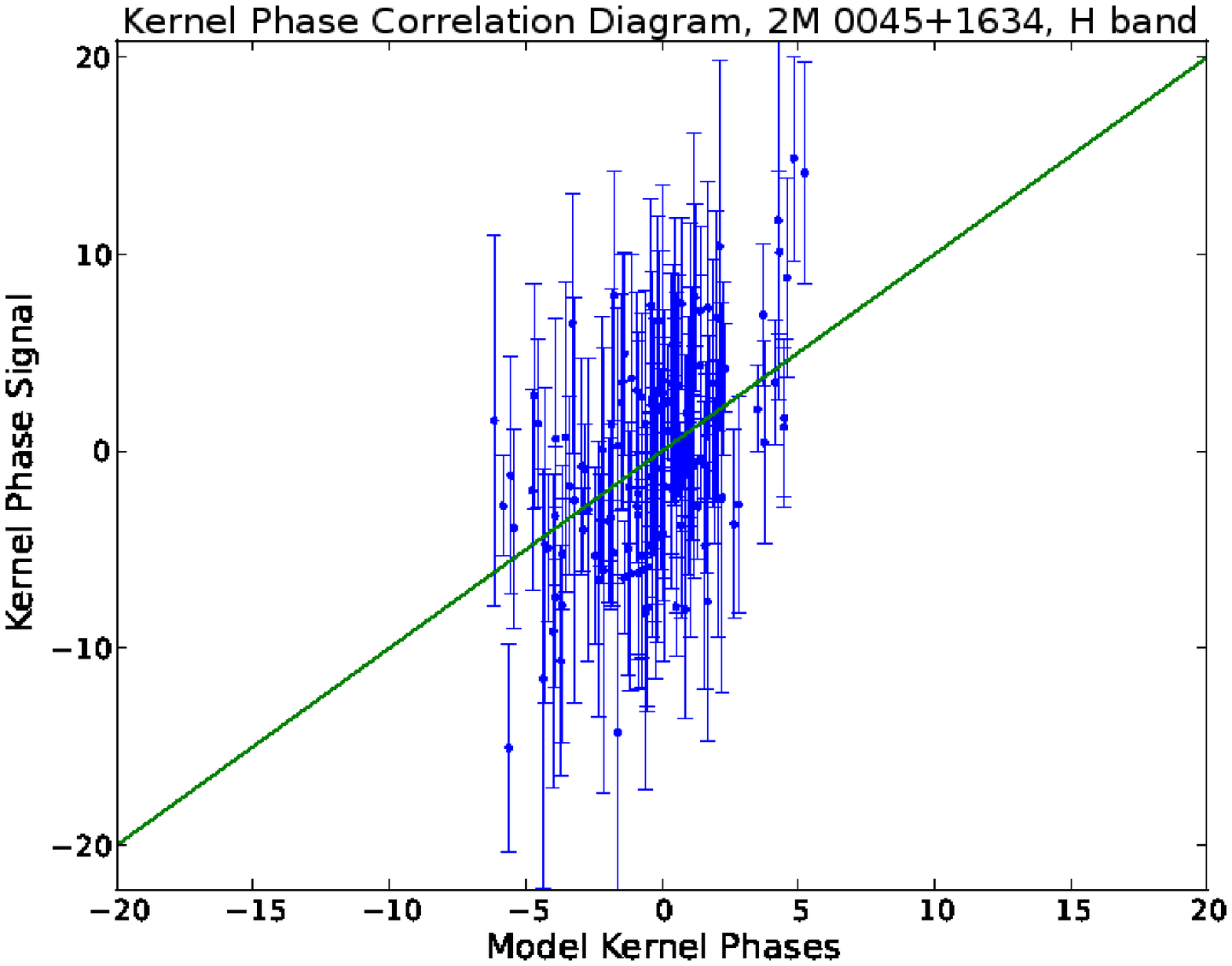}
\caption{Kernel phase correlation diagrams for the newly-detected binary 2M 0045+1634 in (left) $J$ band and (right) $H$ band.}
\label{60corr}
\end{figure*}

\begin{figure*}[h]
\plottwo{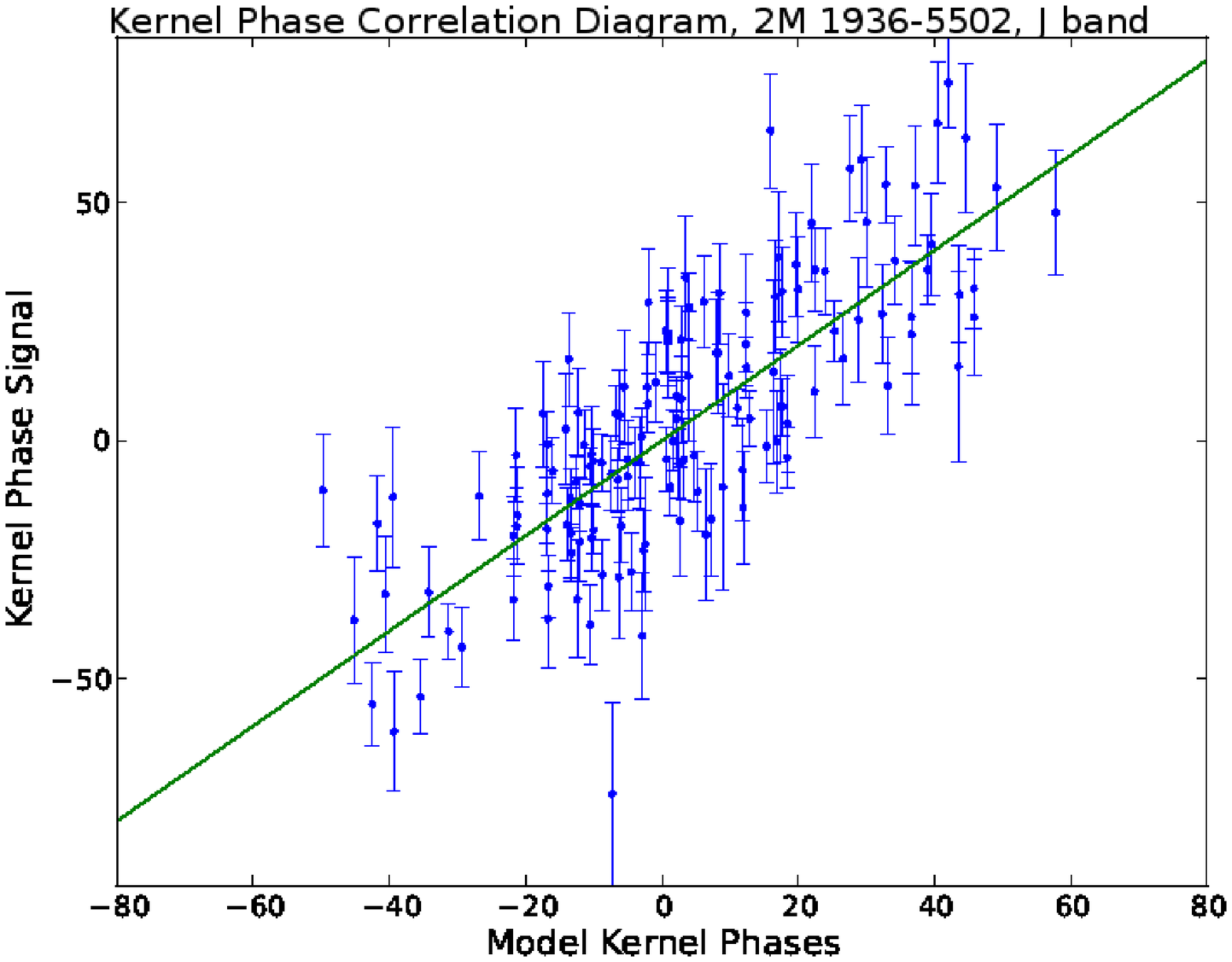}{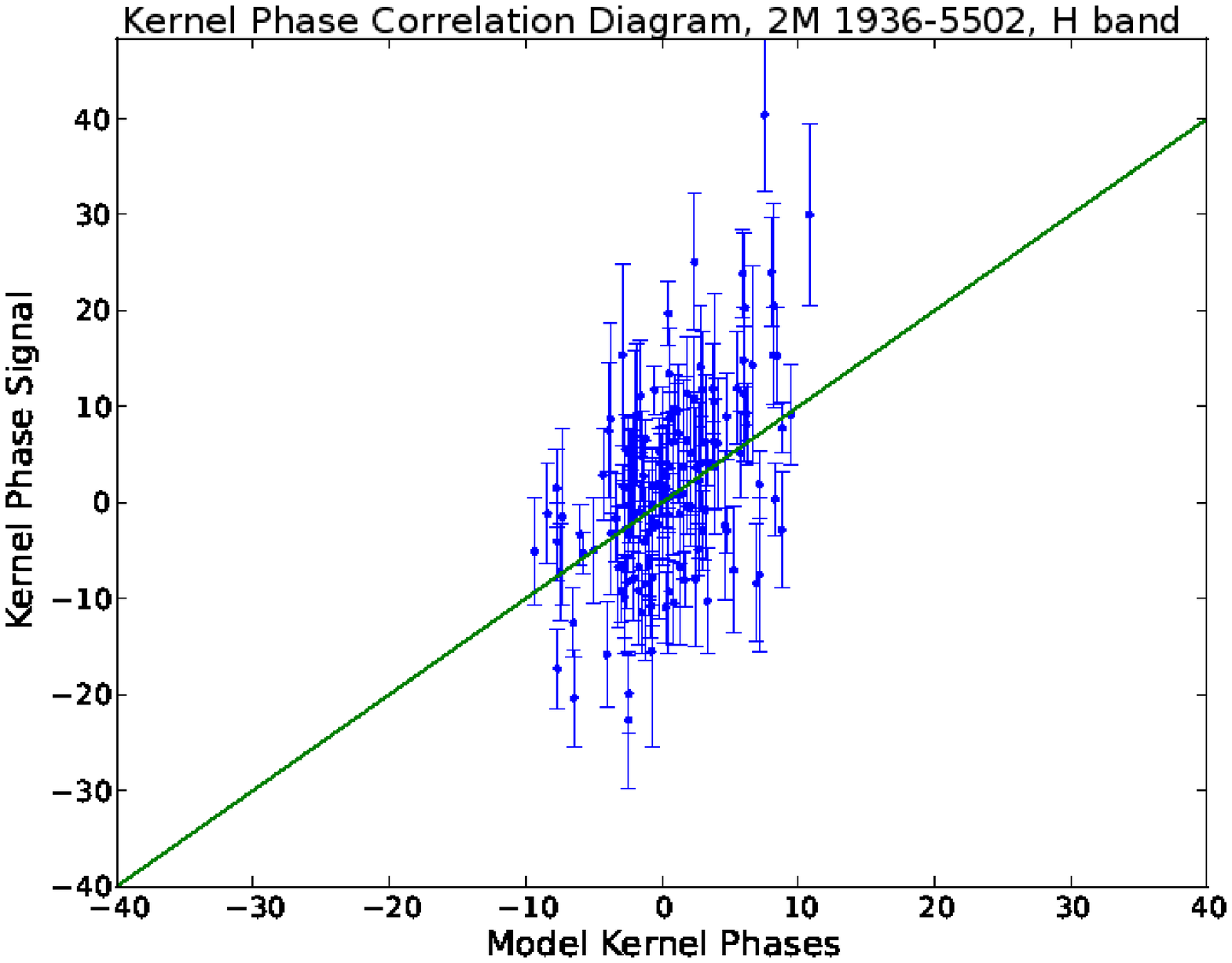}
\caption{Kernel phase correlation diagrams for the newly-detected binary 2M 1936-5502 in (left) $J$ band and (right) $H$ band.}
\label{37corr}
\end{figure*}

\begin{figure*}
\plottwo{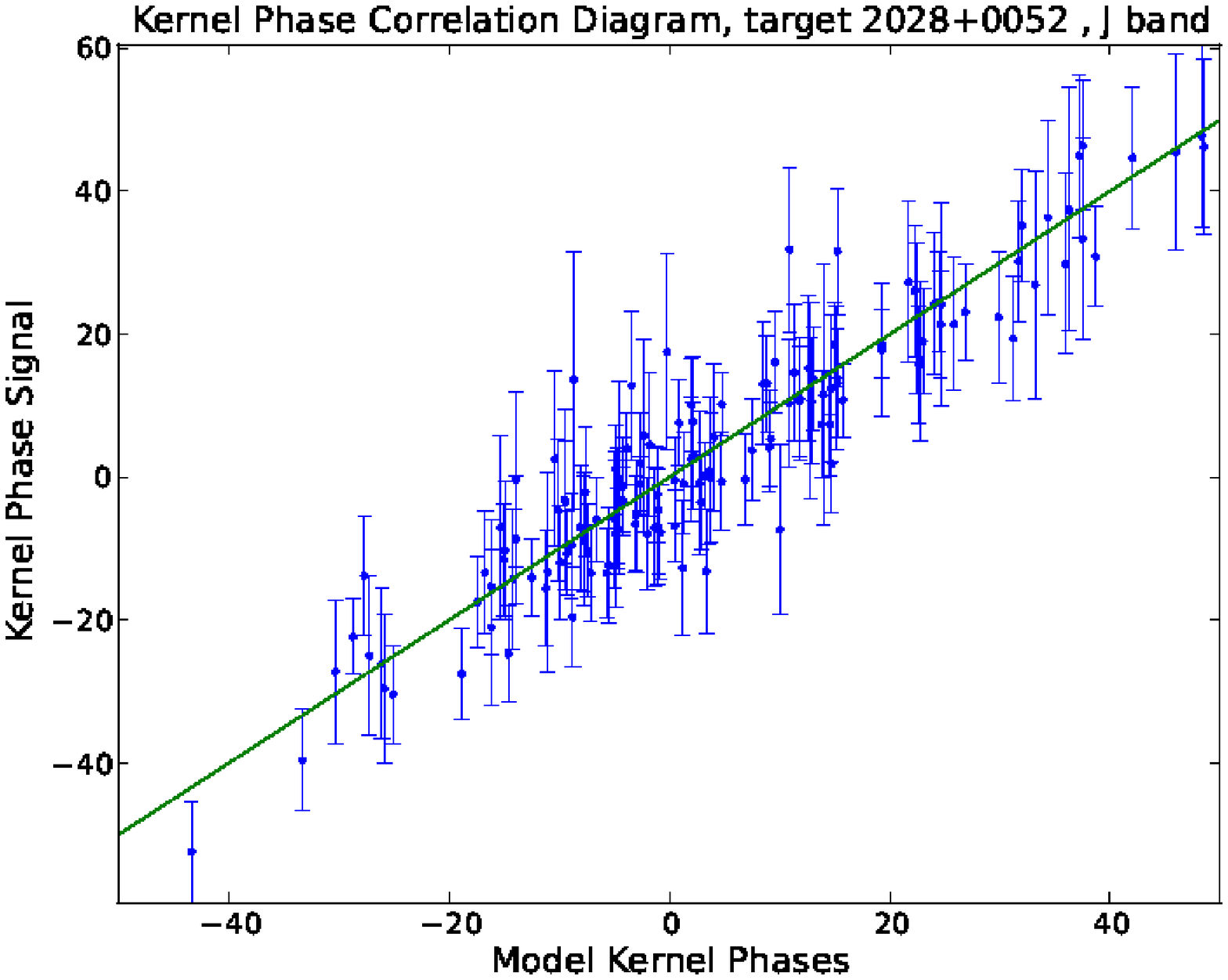}{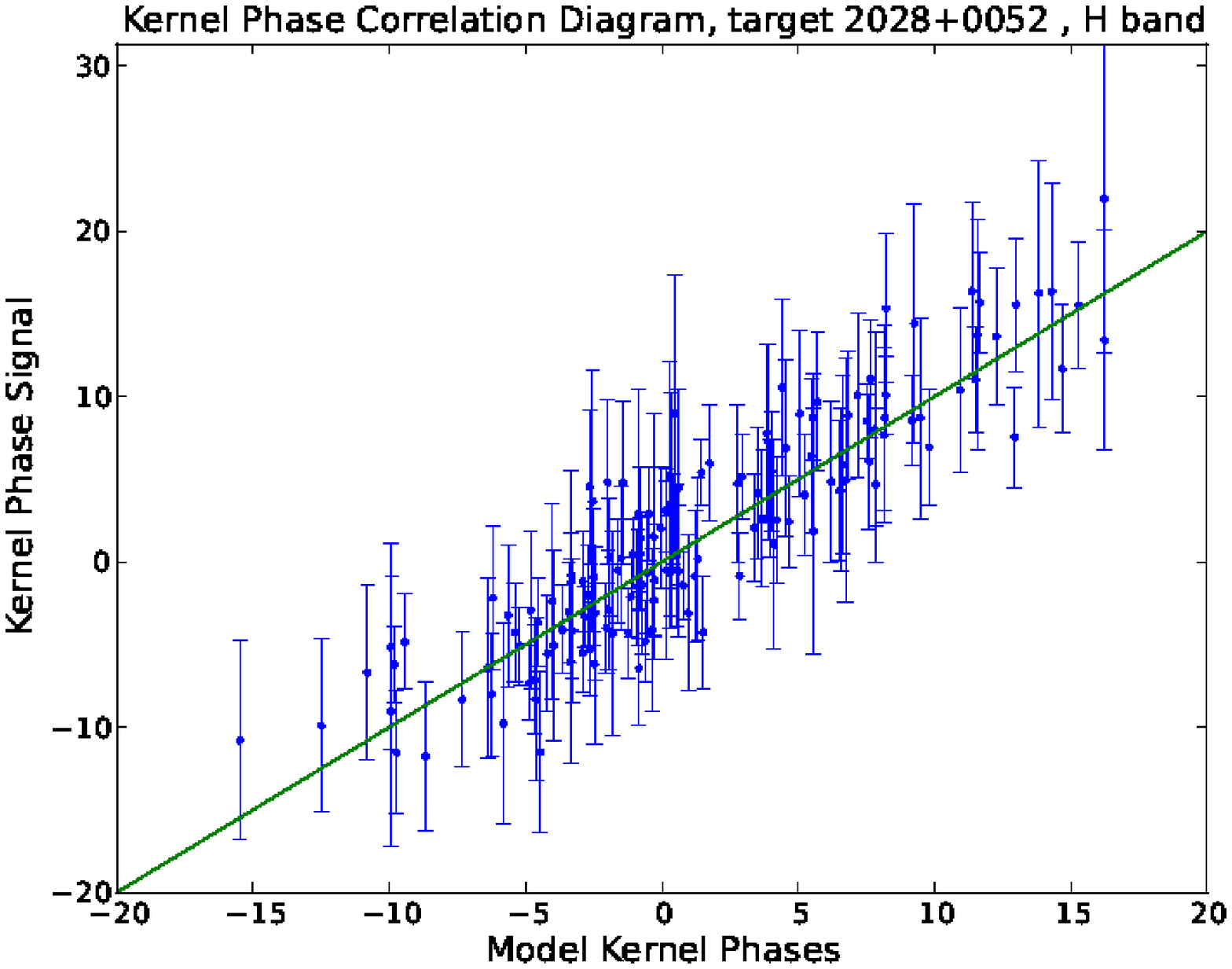}
\caption{Correlation diagrams for the newly-detected binary 2M 2028+0052 in (left) $J$ band and (right) $H$ band.}
\label{21corr}
\end{figure*}

\begin{figure*}
\plottwo{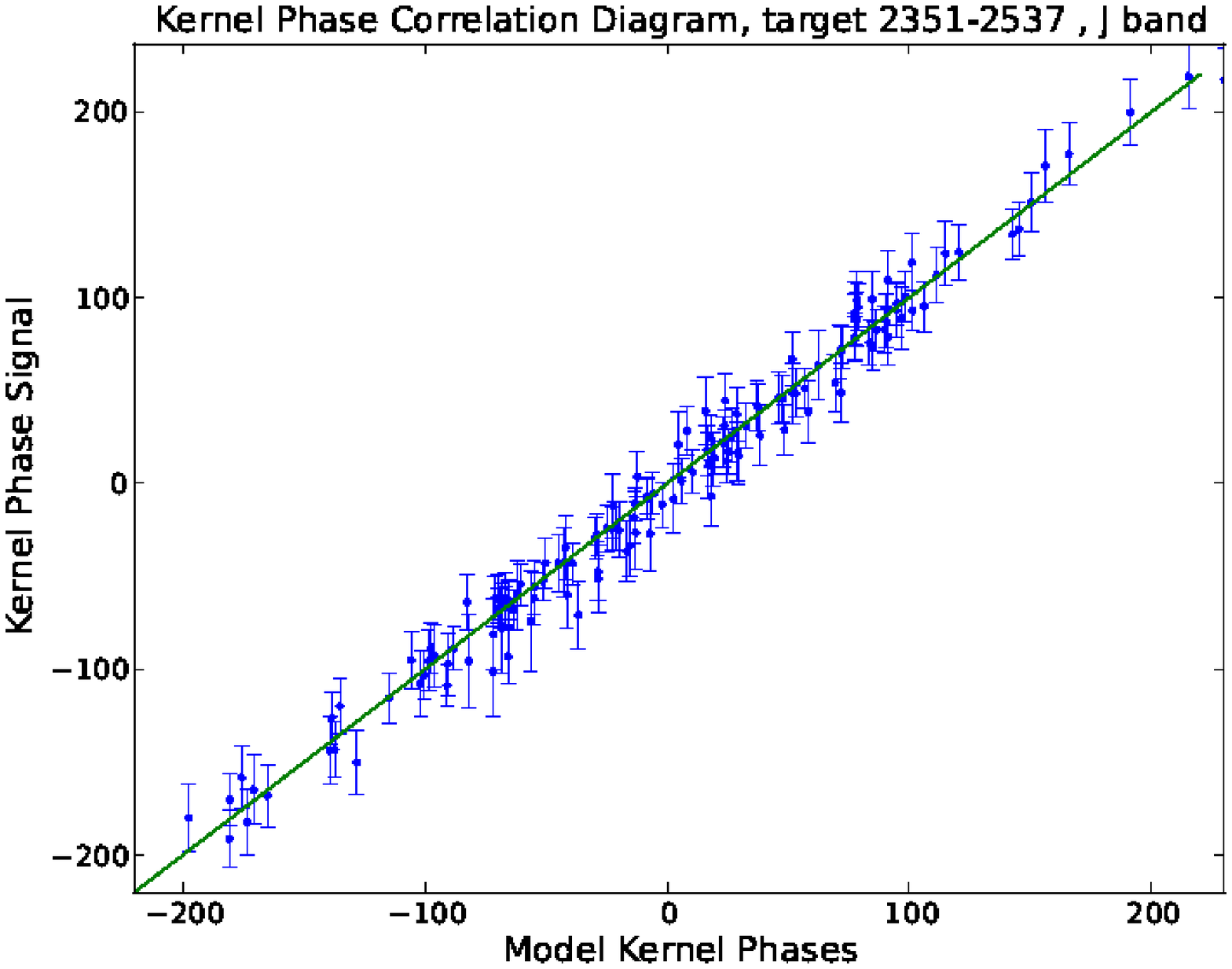}{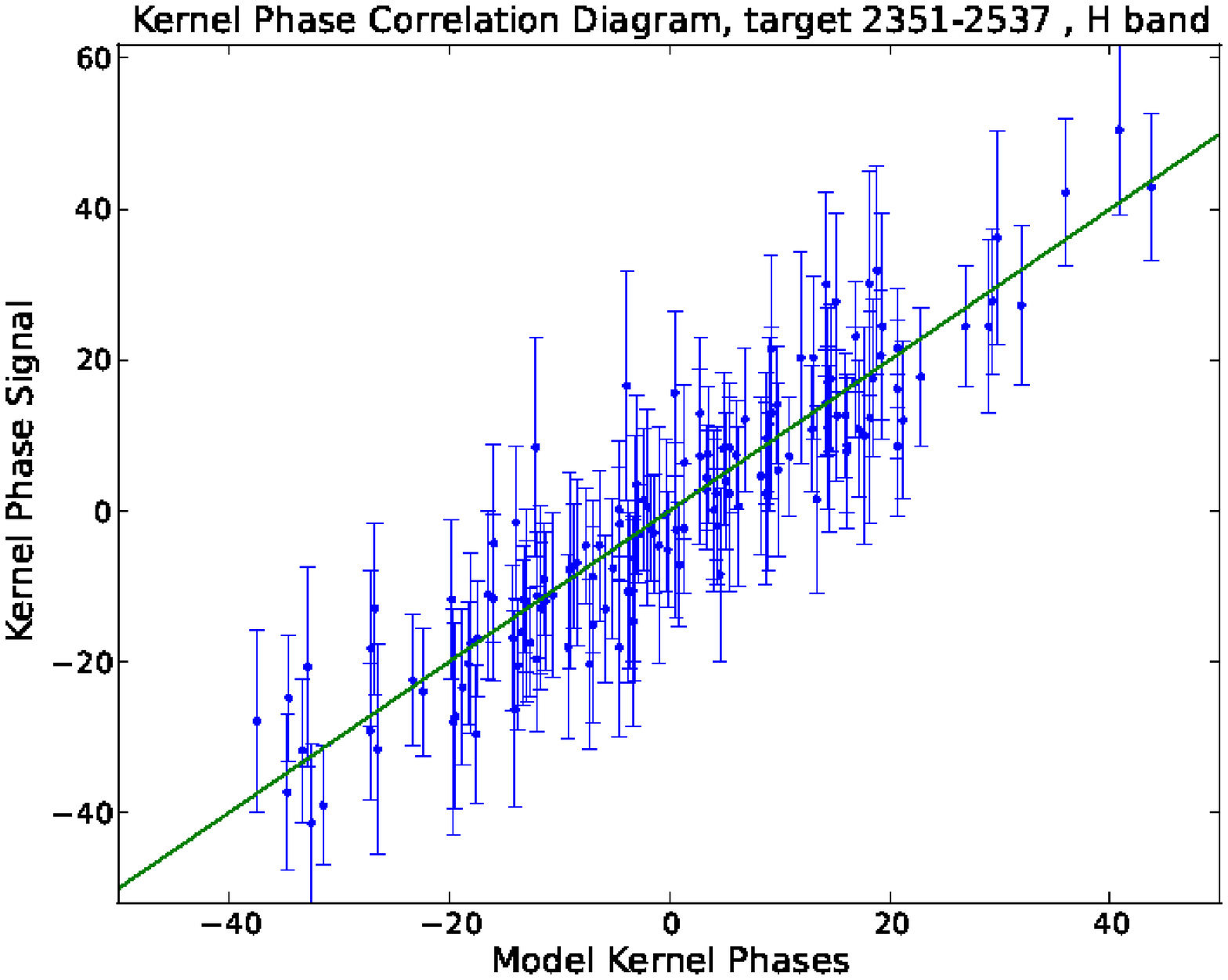}
\caption{Correlation diagrams for the newly-detected binary 2M 2351-2537 in (left) $J$ band and (right) $H$ band.}
\label{25corr}
\end{figure*}

Table~\ref{newbins} reports five firm binary candidates not detected in the original 
\citet{2006AJ....132..891R} or \citet{2008AJ....135..580R} studies, but recovered at
very high 99.9\% confidence from our kernel phase analysis (with the exception of 
2M~0045+1634 for which the detection confidence was only 99\%).
Correlation plots for all of these are displayed as Figures~\ref{60corr} to \ref{25corr}. 

\citet{2006AJ....132..891R} noted three targets which exhibited broad PSFs, however
they went on to report these stars had ``... no evidence for the presence of a secondary 
component, and the broader profiles are probably an instrumental effect.'' These were 2M 1507-1627 and 2M~1936-5502, with a PSF FWHM of 2.47 pixels (106 mas) and 2M~0036+1821 with a FWHM of 2.56 pixels (110 mas), as opposed to the FWHM of 2.3-2.4 pixels found through the rest of the unresolved sample.
Our kernel phase analysis identifies two of these as binary candidates: 2M~1936-5502, and 2M~0036+1821. 
We note that for both of these, the kernel phase signal-to-noise ratio is only of order 2, and 
they exhibit a correspondingly noisy correlation plot. 

2M~1936-5502 supported an alias fit at around $225^{\circ}$ position angle, and the contrast in $H$ band was very poorly constrained. This may well be considered the most marginal fit reported in this section. Nevertheless, both bands support overlapping position angle modes at $330^{\circ}$, and this object was considered for further study.

For 2M~0036+1821, there are two distinct $\chi^2$ minima in $H$ band, one of which overlaps precisely
with the single distinct minimum from the $J$ band data. 
In assigning parameter estimates for this object, we have assumed some uncalibrated source of noise affected
the $H$ band and have therefore restricted the parameter space deliberately to contain only that region 
around the high-significance $J$ band $\chi^2$ detection. 

The third target with a reported broad PSF, 2M~1507-1627, shows marginally-significant companion fits 
with parameters which are inconsistent between $J$ and $H$ bands. 
Until additional data can be recovered with higher signal-to-noise, we classify this object to be
unresolved with no companion. 

A companion to 2M~0036+1821 was also reported by \citet{2010ApJ...715..724B} with 
non-redundant masking interferometry in $K$ band with the laser guide star adaptive optics system on the Hale Telescope at Palomar Observatory.
From data taken in September 2008 (some 3.5 years after the HST observations), these authors 
report a contrast ratio of 13.1 $\pm$ 3.1 and separation of 90 $\pm$ 11\,mas at a position angle of
$114^\circ \pm 5$.
Although changes in the binary separation parameters are to be expected with progress in an
orbit, the markedly different contrast ratio when compared to our fit in Table~\ref{newbins} is, 
at first glance, hard to reconcile.
However we note that both studies employ similar fundamental methodologies and are affected 
by the strong separation-contrast ratio degeneracy previously discussed. 
In particular, the work of \citet{2010ApJ...715..724B} observed only in the $K_s$ band
and therefore enjoyed none of the advantages offered by dual wavelengths in lifting the ambiguity
discussed in Section~\ref{bayes}.
Indeed, in their discussion an alternate detection at 25:1 contrast and 243\,mas with equal probability
is debated and ruled out based on the HST archival data.
To test the hypothesis that both the \citet{2010ApJ...715..724B} companion and the one reported here
are consistent with the same degree of phase asymmetry, we re-fit our kernel phase data with 
an enforced higher contrast ratio above 10:1.
This immediately resulted in much larger best-fit separations, approaching those from \citet{2010ApJ...715..724B}, 
and we therefore conclude that the two studies have probably identified the same companion.

All of these systems were originally assigned spectral classes in \citet{2006AJ....132..891R}, which must be altered to reflect the discovery of new companions. A similar approach is taken here, using tabulated $J-H$ colours \citep{mamajek} to determine new spectral classes, taking into account the contrast ratio found at each wavelength. The class assigned to the secondary from its colour is compared to the expected contrast with the primary as a function of spectral class, and found in each case to be relatively consistent. These classes, being based on contrasts which are themselves subject to error, are accurate only to within one division. Of special interest are 2M~2351-2537 and 2M~1936-5502. In the former case, \citet{2008AJ....135..580R} reassigned the spectral type from L0.5 determined in \citet{2006AJ....132..891R} to M8, and excluded the system from their 20-parsec catalogue of L dwarfs on these grounds. On the other hand, \citet{2011AJ....141...54A} classify it as L0.0 based on precision photometry. Revealing it as a binary system, its colours and luminosity imply that both primary and secondary must be early-L dwarfs, with possible classes L0 and L1.

Given the uncertainty in contrast in the 2M~1936-5502 system, no accurate spectral type can be determined for the secondary component. If the high-contrast fit is confirmed, then the secondary must have a spectral type later than T9. A brown dwarf of Y class is possible in principle, but coeval companionship to an early-L primary seems unlikely if the primary is a brown dwarf. If this is the case, then the companion may be of planetary mass. On the other hand, if the primary is a star, then this could be an evolved companion of brown dwarf mass as discussed in \citet{2012ApJ...753..156K}. We note that \citet{2012ApJ...752...56F} present a new parallax distance for this object of $15.08\pm1.2$ pc, which compares well with the spectroscopic distance estimate in \citet{2006AJ....132..891R} and comparatiely poorly with new estimates we calculate with the revised tables of \citet{mamajek}, from which we obtain a distance of $18.6\pm2.8$ pc, though we note that the measurement is still consistent with the quoted uncertainties. 

For 2M~0036+1821, the distance is known from trigonometric parallax to be $8.77\pm0.06$ pc \citep{2002AJ....124.1170D}. The angular separation of 44.5~mas therefore equates to 0.4~AU projected physical separation. This is therefore one of the closest projected separations of any resolved brown dwarf binary. 
The 90\,mas separation from \citet{2010ApJ...715..724B} gives typical binary orbital periods of $\sim$2--3\,years, while on the other hand our 44.5\,mas separation gives orbits of $\lesssim$\,1\,year.
Ignoring the changes in separation (due to degeneracy errors), the position angle change of 84$^\circ$ observed between
the HST and Palomar datasets could be consistent with more than one orbit in the former case or several orbits in 
the latter: either is consistent with the data and a more rapid observing cadence is required to unambiguously 
follow the orbit in this particular system.

Excepting 2M~0036+1821 and 2M~1936-5502, the distances of other objects are known from spectroscopic parallax and are therefore subject to change depending on whether the system is found to be a binary. In general, discovery of a companion implies a higher total luminosity for the system, and therefore a greater distance. Table~\ref{newbins} lists recalculated spectroscopic distances, with standard $15\%$ uncertainties. In particular, the reassignment of 2M~2351-2537 as an L0/L1 binary moves it back into the 20-parsec L dwarf sample, while 2M~0045+1634 and 2M~2028+0052 fall beyond 20 parsecs. The spectroscopic distance of $9.2\pm 1.4$ pc for 2M~0036+1821 agrees well with the spectroscopic parallax of 8.77 pc.

Our kernel phase re-analysis of the HST archive does not detect a binary reported by \citet{2010ApJ...715..724B} in 2M~0355+1133 at 90 mas separation. 
One explanation may be that at the HST epoch the projected separation was smaller and therefore unresolved. This is nevertheless included in the discussion of binarity fraction in Section~\ref{binarity}.
The object 2M~0045+1634 was observed by \citet{2010ApJ...724....1S} with differential spectral imaging using the HST NICMOS camera; a PSF variation was noted but no other signal found in the vicinity of the brown dwarf.


Thus four of the five detections reported in Table~\ref{newbins} can be considered new, and we confirm
the previously-reported detection in 2M~0036+1821 albeit with discrepant best-fit contrast.
All detections are significantly below the formal diffraction limit and at separations of the order 
of the detector plate scale, which is 43.1 mas/pixel. 
It is therefore unsurprising that the candidate binaries are a subject to parameter correlation and 
substantial uncertainties. 
These may therefore require confirmation and accurate constraint of parameters found with follow-up 
observations employing a larger telescope. 

\begin{deluxetable*}{lcccccccc}
\tablecolumns{9}
\tabletypesize{\small}
\tablecaption{Model Parameters for New L-Dwarf Binary Candidates}
\tablewidth{0pt}
\tablehead{
  \colhead{2 MASS} & \colhead{Spectral} & \colhead{Distance} &\colhead{Sep.} & 
  \colhead{Pos. angle} & \colhead{Contrast} & \colhead{Contrast}  &\colhead{J} & 
  \colhead{H} \\
  \colhead{Number} & {Type} & \colhead{(pc)} &\colhead{(mas)} & 
  \colhead{(degrees)} & \colhead{Ratio ($J$)} & \colhead{Ratio ($H$)} &\colhead{mag} & \colhead{mag} 
}
\startdata
0036+1821\tablenotemark{a}\tablenotemark{b} & L3.5  & 8.77 $\pm$ 0.06 \tablenotemark{d}& 44.5$\pm$1.2 & 198.4$\pm$1.3 & 1.85$\pm$0.3  & 2.9$\pm$1.1 &  12.47 & 11.59  \\
......... A  & L4 &&&&&     &  12.93 & 11.91  \\
......... B  & L5-6 &&&&&    &  13.6  & 13.06 \\
0045+1634\tablenotemark{c}   & L0   &26.8 $\pm$ 4.0 & 50.3$\pm$0.7 & 300.6$\pm$2.5 & 1.11$\pm$0.02 & 1.12$\pm$0.03 &  13.06 & 12.06   \\ 
......... A  & L0 &&&&&     &  13.75 & 12.75 \\%
......... B  & L0 &&&&&     &  13.87 & 12.87  \\%
1936-5502\tablenotemark{a}  & L4    &$15.08\pm 1.2$\tablenotemark{e} & 67.1$\pm$6.4 & 330.9$\pm$1.0 & 17.7$\pm$3.9 & 35.5$\pm$8.1 &  14.49 & 13.63  \\ 
......... A  & L4 &&&&&     &  14.54 &  13.66 \\
......... B  & (T-Y) &&&&&     &  17.67 & 17.53  \\
2028+0052                   & L3   &26.1 $\pm$ 3.9    & 45.8$\pm$1.2 & 107.7$\pm$1.1 & 1.52$\pm$0.1 & 3.1$\pm$ 0.5   &  14.3  & 12.38 \\
......... A  & L3 &&&&&     &  14.85 & 12.7  \\
......... B  & L4 &&&&     &  15.3  & 13.9  \\
2351-2537    & L0.5      &     17.8 $\pm$   2.7   & 63.3$\pm$0.3 & 348.8$\pm$0.3 & 2.4$\pm$0.1 & 2.24$\pm$0.2 &  12.46 & 11.73  \\ 
......... A  & L0  &&&&&     &  12.84 & 12.13  \\
......... B  & L1 &&&&&     &  13.79 & 13.006 \\
\enddata
\label{newbins}
\tablenotetext{a}{\citet{2006AJ....132..891R} listed these objects as having a significant PSF abnormality. A third object showing such an abnormality, 2M 1507-1627, does not show a kernel phase binary fit.}
\tablenotetext{b}{\citet{2010ApJ...715..724B} detected this binary with Palomar adaptive optics aperture masking.}
\tablenotetext{c}{This object was observed with spectral differential imaging by \citet{2010ApJ...724....1S}. A PSF broadening was noticed but no other unambiguous signal was detected.}
\tablenotetext{d}{This distance is known from trigonometric parallax to be $8.77\pm0.06$ pc \citep{2002AJ....124.1170D}.}
\tablenotetext{e}{Using the newer spectral class - absolute magnitude tables from \citet{mamajek}, we revise the Reid spectroscopic distance from 15.4 pc to $18.6\pm2.8$ pc.}
\end{deluxetable*}

\begin{deluxetable}{lccc}
\tablecolumns{4}
\tabletypesize{\small}
\tablecaption{Mass Estimates for  New L-Dwarf Binary Candidates}
\tablewidth{0pt}
\tablehead{
  \colhead{2 MASS} & \colhead{$M_{0.5 Gyr}$} &\colhead{$M_{1 Gyr}$} & 
  \colhead{$M_{5 Gyr}$}\\
  \colhead{Number} & \colhead{$M_{\astrosun}$}& \colhead{$M_{\astrosun}$}
& \colhead{$M_{\astrosun}$}
  }
\startdata
0036+1821 A  &  0.049  &  0.064 & 0.073  \\
......... B  &  0.045  &  0.060 & 0.071 \\
0045+1634 A  &  0.07   &  0.078 & 0.081 \\%
......... B  &  0.07   &  0.078 & 0.081 \\%
1936-5502 A  &  0.049  &  0.064 & 0.073  \\
......... B  &         &        &   \\
2028+0052 A  &  0.052  &  0.066 & 0.074  \\
......... B  &  0.049  &  0.064 & 0.073  \\
2351-2537 A  &  0.07   &  0.078 & 0.081 \\%
......... B  &  0.059   &  0.072 & 0.075 \\
\enddata 
\label{newbinsmass}
\end{deluxetable}

\subsection{Possible New Companions at Higher Contrasts}
\label{highcon}

Estimates in \citet{2011SPIE.8151E..33M} suggested the possibility of detections at contrast ratios greater than 50:1 using kernel phases. The recent discovery of planetary-mass companions to brown dwarfs \citep{2010ApJ...714L..84T} indicates that these objects exist and are at least common enough that in a sufficiently large dataset candidates may be found. On the other hand, given the parameter degeneracy, it is possible that some high contrast detections revealed by these techniques are in fact very close low contrast binaries. Owing to this large uncertainty, no attempt has been made to assign separate magnitudes or spectral classes to the faint companions identified in this section.

A study of all the unresolved objects in our sample extending to a contrast of 100 revealed four objects with fits agreeing in position angle between bands: 2M 0314+1603, 2M 1539-0520, 2M 0830+4828 and 2M 0109+2949. 

These objects were then subjected to a nested sampling joint fit to examine correlation at best fit. In the extreme contrast-separation regime probed here, the parameters show very pronounced degeneracy and it is impossible to reliably distinguish between high contrast objects further out and lower contrast objects closer in, even using both $J$ and $H$ bands. No error was added in quadrature, as in each case the kernel phases were already over-fit with reduced $\chi^2 < 1$. 

2M 0314+1603 is the best of the candidates in Table~\ref{margtable} and exhibits excellent correlation in $J$ band. Nevertheless, while $H$ band data favours a $\chi^2$ minimum at this same position angle, it is unconstrained in contrast ratio and reliably runs off to high contrast in each attempted fit. Likewise 2M 1539-0520 supported good correlation diagrams but a surprisingly large discrepancy in contrast ratio between bands. This is likely to be the result of a strongly degenerate fit which constrains neither contrast well; the contrast ratios should therefore not be considered well-determined.

With a $J$ contrast $>160$, as a coeval companion to a brown dwarf this can only be a planetary mass object. On the other hand, it is possible that the L primary may be a star at the low-mass limit of the main sequence and its companion is itself an evolved brown dwarf. Such a system is still of considerable interest: as noted in \citet{2012AJ....144...64D}, low-mass binary systems tend to be of equal mass, increasingly so at lower primary masses. Accordingly, few brown dwarf secondaries to main sequence primaries are known, especially at close separations. This is the low-mass end of the `brown dwarf desert' first identified by \citet{2000PASP..112..137M}, which has since become the subject of intense study \citep{2006ApJ...640.1051G,2007AJ....133..971A,2008ApJ...679..762K,2008A&A...491..889D,2011ApJ...731....8K,2012AJ....144...64D,2012ApJ...744..120E}. A confirmation of the status of this system, regardless of result, would therefore be a potentially significant finding.

\begin{figure*}[h]
\plottwo{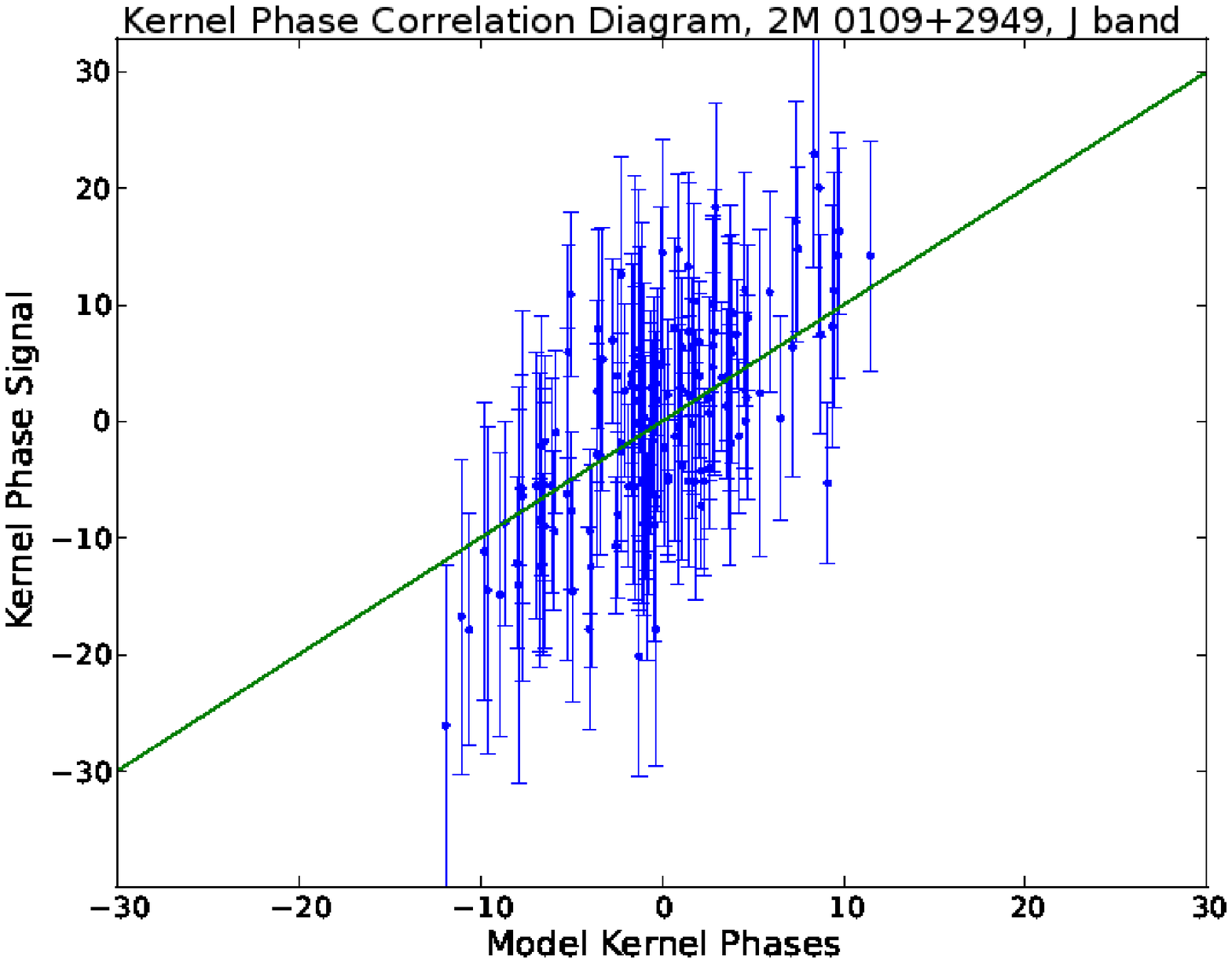}{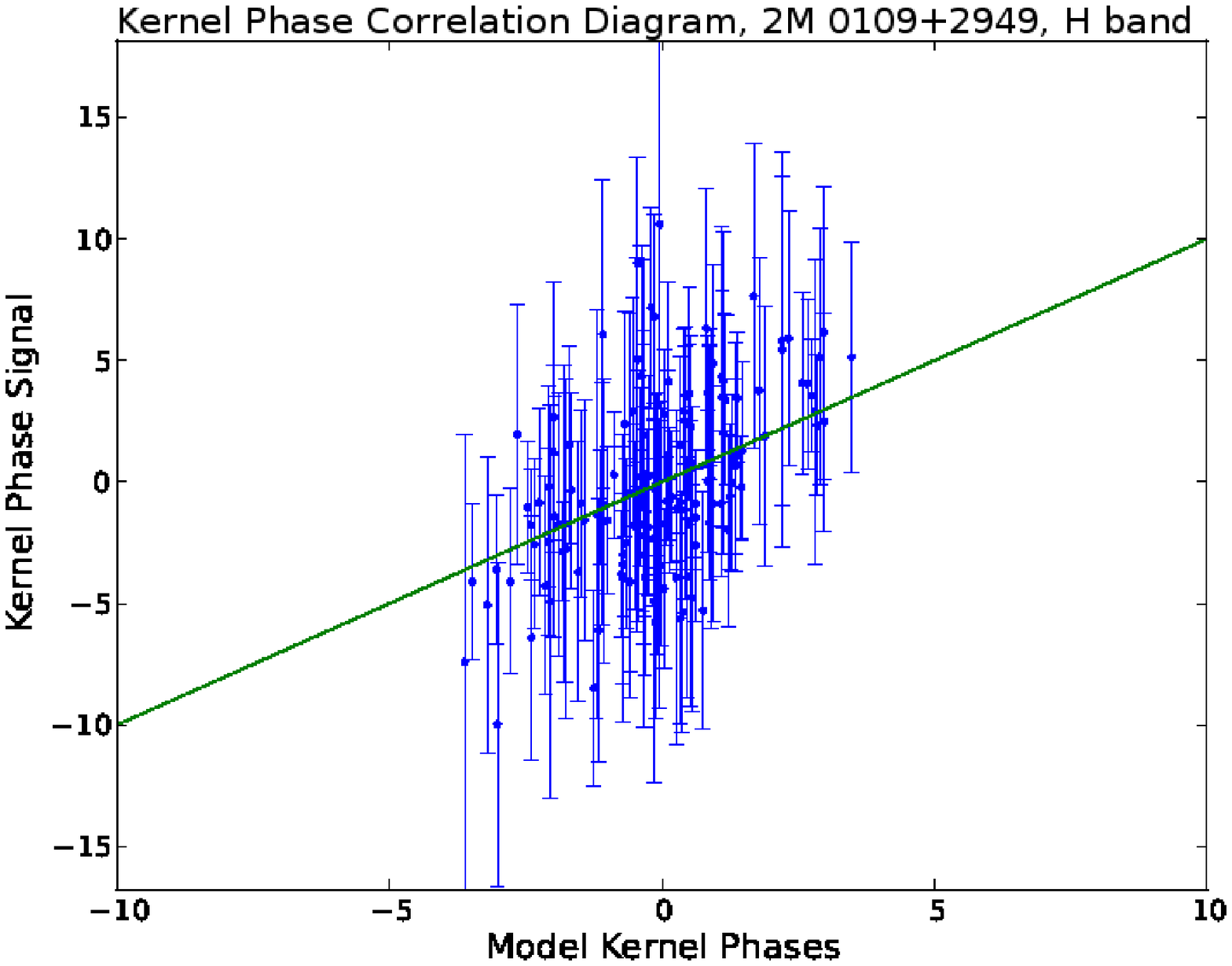}
\caption{Kernel phase correlation diagrams for marginal detection 2M 0109+2949 in (left) $J$ band and (right) $H$ band.}
\label{23corr}
\end{figure*}

\begin{figure*}[h]
\plottwo{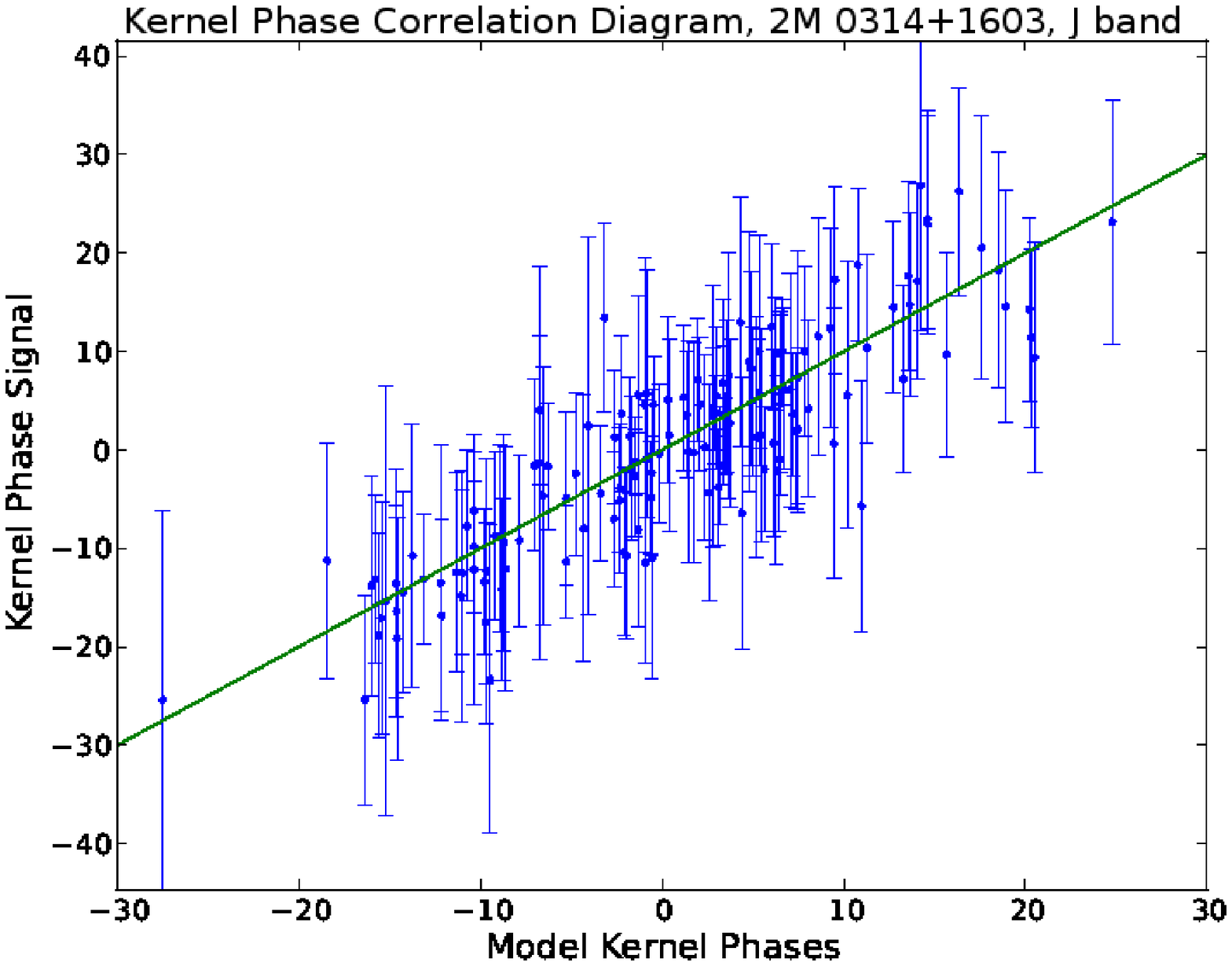}{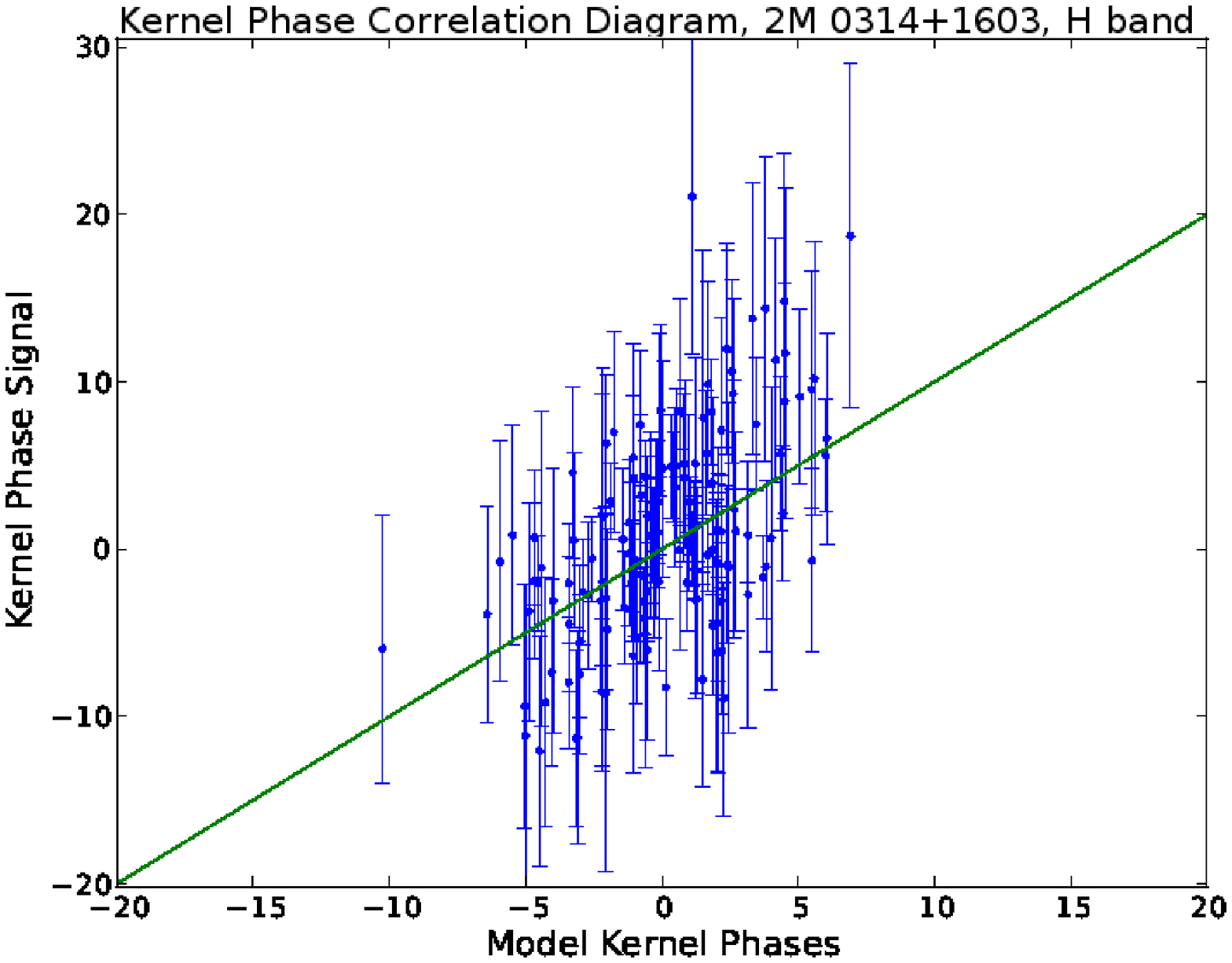}
\caption{Kernel phase correlation diagrams for marginal detection 2M 0314+1603 in (left) $J$ band and (right) $H$ band.}
\label{31corr}
\end{figure*}

\begin{figure*}[h]
\plottwo{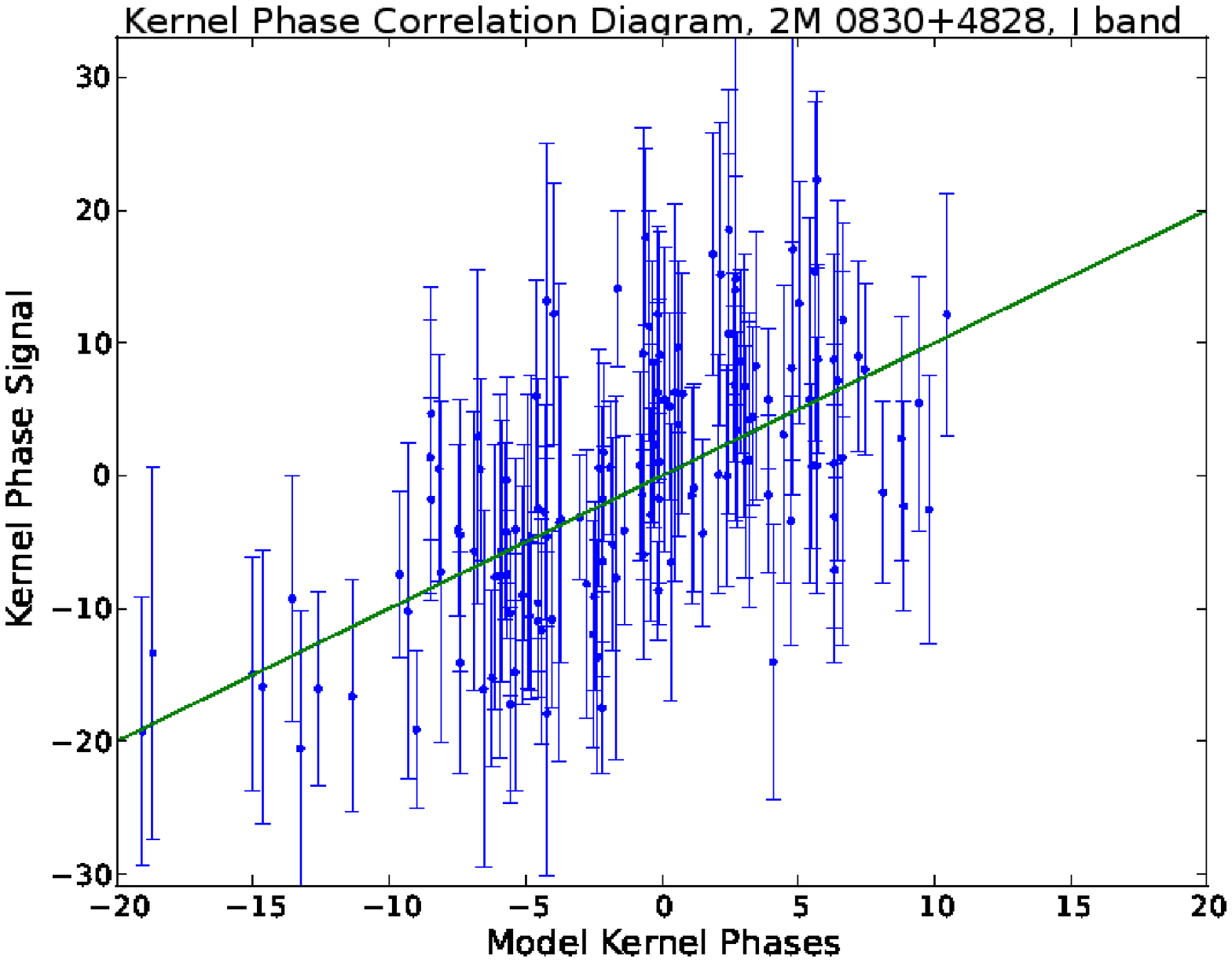}{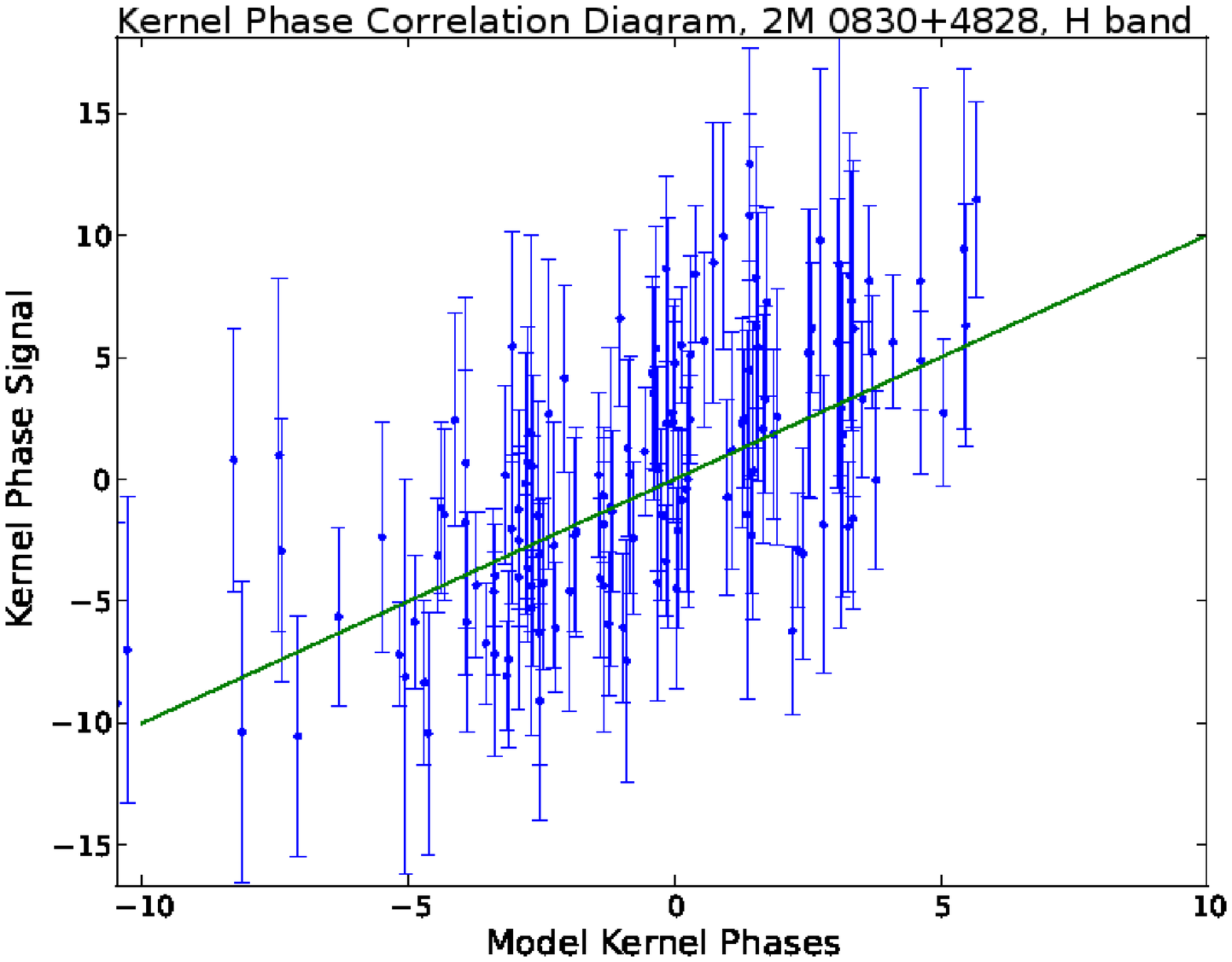}
\caption{Kernel phase correlation diagrams for marginal detection 2M 0830+4828 in (left) $J$ band and (right) $H$ band.}
\label{03corr}
\end{figure*}

\begin{figure*}[h]
\plottwo{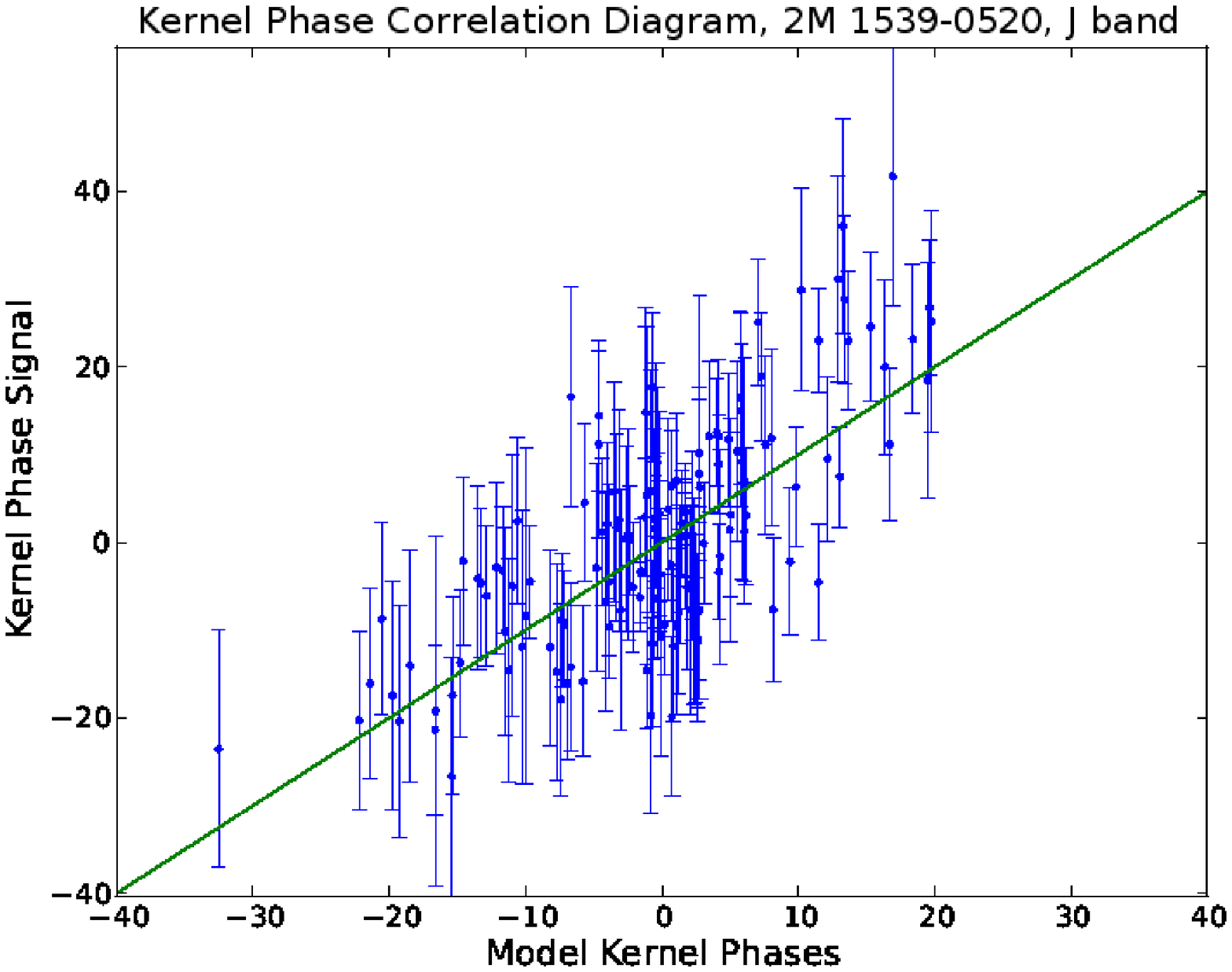}{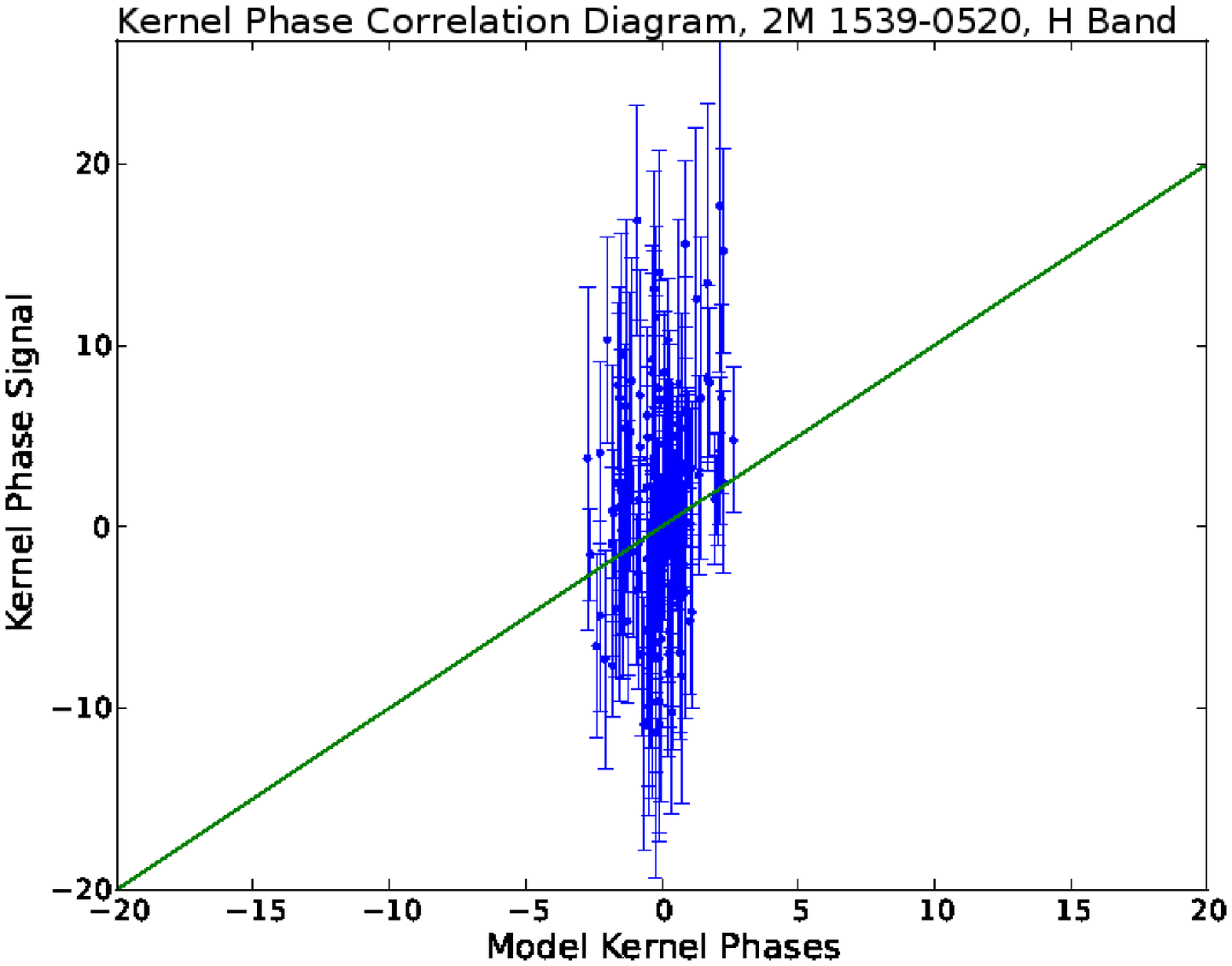}
\caption{Kernel phase correlation diagrams for marginal detection 2M 1539-0520 in (left) $J$ band and (right) $H$ band.}
\label{44corr}
\end{figure*}

Correlation plots for the marginal detections are shown in Figures~\ref{23corr} to \ref{44corr}. 

\begin{deluxetable}{lcccc}
\tablecolumns{5}
\tabletypesize{\small}
\tablecaption{Model Parameters for Marginal High Contrast Companions}
\tablewidth{0pt}
\tablehead{
  \colhead{2 MASS} &\colhead{Separation} & 
  \colhead{Position angle} & \colhead{Contrast} & \colhead{Contrast} \\
  \colhead{Number} &\colhead{(mas)} & 
  \colhead{(degrees)} & \colhead{Ratio ($J$)} & \colhead{Ratio ($H$)}
}
\startdata
0109+2949  & 49 $\pm$ 13 & 268 $\pm$ 3 & 40 $\pm$ 25 & 43 $\pm$ 29   \\ 
0314+1603  & 124 $\pm$ 16 & 227 $\pm$ 2 & 70 $\pm$ 10  & $>$ 160 \\ 
0830+4828  & 48 $\pm$ 9 & 120 $\pm$ 3 & 29 $\pm$ 16 & 14 $\pm$ 11 \\ 
1539-0520  & 35 $\pm$ 5 & 332 $\pm$ 5 & (4 $\pm$ 3)  & (28 $\pm$ 16) \\ 
\enddata
\label{margtable}
\end{deluxetable}

\subsection{Survey Confidence and Detection Limits}
\label{deteclimits}

\begin{figure}[h]
\center
\includegraphics[width=0.5\textwidth]{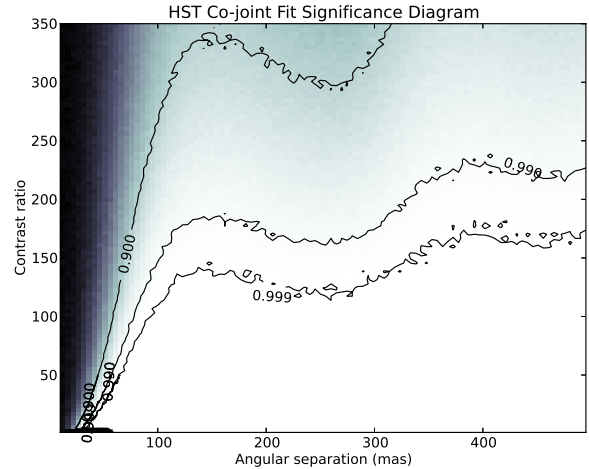}
\caption{HST snapshot contrast-detection limits for  co-joint fitting as a function of separation, averaged over position angle. The $90\%$, $99\%$ and $99.9\%$ contours are overplotted with labels. Paler regions indicate higher significance.}
\label{cojointsig}
\end{figure}

\begin{figure}[h]
\center
\includegraphics[width=0.5\textwidth]{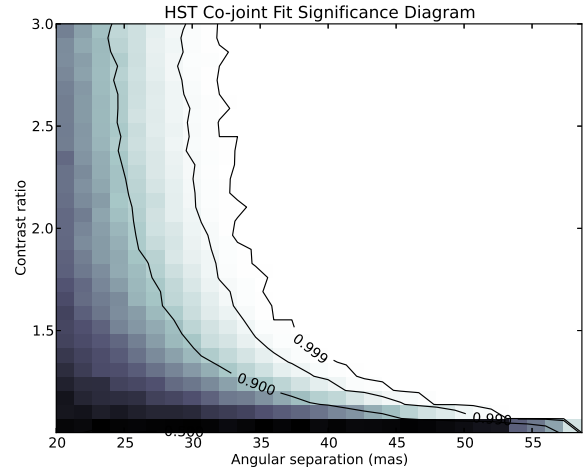}
\caption{HST snapshot contrast-detection limits for  co-joint fitting as a function of separation, averaged over position angle. The $90\%$, $99\%$ and $99.9\%$ contours are overplotted labels. Paler regions indicate higher significance.
This figure represents a simulation of the region near the origin of Figure~\ref{cojointsig} with a finer sample grid. Note the turnaround at low contrast: kernel phase performs poorly at detecting very-low-contrast companions.}
\label{cojointsignear}
\end{figure}
\vspace{0cm}

In order to quantify our survey detection threshold as a function of model parameters, 
we performed a Monte-Carlo study simulating detection of a population of model binaries.
We ran 100 simulations, adding a binary at each point on a grid in separation, 
position angle and contrast ratio, in both $J$ and $H$ bands, as a co-joint fit. 
We then added Gaussian noise randomly to each of these, with the distribution given by our 
measured error distribution as discussed in Section~\ref{sec:kpanalysis}. 
The detection rates as a function of separation and contrast are averaged over position angle
(distributions were found to be azimuthally symmetric).
Quantitative detection thresholds were formulated from comparison of the binary model $\chi^2$ 
to the null hypothesis. 
This method is similar to that used in \citet{2010ApJ...724..464M}. 
We also examined the low-contrast, low separation domain in which we have found many of 
our binaries with a finer sample grid. The contrast detection limit curves turn around at
very low contrasts, as kernel phase performs poorly in discerning very-low-contrast companions,
which give rise to highly-symmetrical images with a weak Fourier phase signal.
This is true of any phase-based method, as the Fourier phases encode information about
spatial asymmetries in a source image, and this problem therefore affects closure phases
in NRM interferometry as well. This therefore places a lower limit on the contrasts detectable
at very small separations.
The contrast thresholds obtained from these studies are shown in Figures~\ref{cojointsig} 
and \ref{cojointsignear} respectively.
Our estimates here compare favourably with the kernel phase contrast detection limits predicted for the Keck Telescopes in \citet{2011SPIE.8151E..33M}, which are at best closer to 50:1. This very high sensitivity with the HST is unsurprising, given the high quality of the wavefront, and this technique therefore holds great promise in its application to similar snapshot samples.

Over any companion detection hangs the question as to whether the pair of stars imaged are physically associated, or whether they merely happen to lie along the same line of sight but are otherwise unrelated. It is therefore important to establish the expected count of background stars in the direction of each star, which will in general differ according to the star's position, owing to shape of the Galaxy and the distribution of dust. 

The \emph{Galaxia} software package \citep{2011ApJ...730....3S} is a synthetic survey tool, which calculates the expected density of Galactic stars at a given magnitude in a given band along a given line of sight. This was applied in 0.1-square-degree regions around the coordinates of each binary candidate, searching for background stars between 12th and 18th apparent magnitude in $J$ and $H$ bands. For a canonical 12th magnitude primary, these span the contrast range from 1 to 100, and therefore cover both the brown dwarf and planetary-mass companion regimes. This was then scaled down to the expected counts in a 200 mas circle around the target.

In summary, no field in either band shows significantly more than 2$\%$ probability of finding a background star within 200 mas, and in the overwhelming majority of cases, including almost all new binary candidates and marginal detections, this figure was an order of magnitude lower. It is therefore exceedingly likely that background stars do not contribute in any way to the population of companion candidates in this survey.
\section{Discussion}
\label{sec:disc}

\subsection{Kernel Phase Performance}
\label{perf}

From our retrieval of existing and new binaries, we have demonstrated that kernel phase
interferometry performs well in medium and wide band filters.
For shorter wavelengths the PSF quality degrades increasing the errors, however in all
cases space telescope data are firmly in the regime where wavefront quality is excellent
and easily sufficient for the purposes of the algorithm.
This method has demonstrated the delivery of very precise astrometry for medium-separation
systems.
If employed over multiple epochs, this will permit correspondingly accurate dynamical mass
measurements for most detected binaries.

For the very closest companions, we note that the signal-to-noise ratio is lower and
accordingly kernel phase yields weaker constraints on the binary parameters, which
nevertheless should still permit the determination of dynamical masses.
For systems whose close separation puts them beyond the diffraction limit, errors are dominated by covariance between
the separation and contrast creating model ambiguity between close, bright companions and
distant, fainter companions.
We have shown that when multi-wavelength observations are employed, this degeneracy
can be partially lifted.

\subsection{Opportunities for Dynamical Mass and Radius Measurement}
\label{opportunity}

Adopting distance and mass estimates from \citet{2006AJ....132..891R} and \citet{2008AJ....135..580R}, 
and our own parameter estimates, typical binary objects reported here such as 2M~2351-2537 and 2M~2028+0052 
would have a binary separation of $\sim$ 1 AU and therefore orbital periods of 3 to 4 years. 
A follow-up campaign with LGS AO could be used to track the orbits of these binaries and dynamically determine 
their mass in the near term, as has already been achieved for GJ~802B in \citet{2008ApJ...678..463I}. 

With the already-observed epochs from \citet{2006AJ....132..891R,2008AJ....135..580R} and \citet{2010ApJ...715..724B}, two more epochs should be sufficient to permit a first fit to the binary parameters of each system. Targets observed in 2006 or 2008 will have completed a substantial fraction of their orbit since they were first observed; on the other hand, over a two month period they will have rotated by 6 degrees in mean anomaly, which is substantially greater than the $\sim$ degree uncertainties in position angle obtained with kernel phase or aperture masking. These are pessimistic figures, in that the systems are early L dwarfs and include several systems substantially closer than 2 AU, which leads to correspondingly shorter orbital periods for targets of interest.

In Table~\ref{newbinsmass}, we present approximate masses computed using the methods of \citet{2006AJ....132..891R}. These were calculated by taking the absolute $J$ magnitude of the new spectral class as listed in \citet{mamajek}, applying $J$ magnitude bolometric corrections from \citet{2004AJ....127.3516G} and comparing bolometric magnitudes with the 0.5, 1 and 5 Gyr isochrones from \citet{2000ApJ...542..464C}. The calculated values are subject to direct uncertainties of order $\sim 0.03 M_{\astrosun}$: despite the significant uncertainty in photometry, bolometric magnitude is a very strong function of mass and age and purely photometric uncertainties are small. On the other hand, the models used come with significant caveats, especially because brown dwarfs cool as they age: the mass estimates, given photometry, therefore depend on the assumed age of the system. Obtaining dynamical masses is key to calibrating these models brown dwarfs in general, as discussed in Section~\ref{intro}. These close new binaries therefore present a significant addition to the population of targets which can be followed up on short time scales.

In addition to this, many brown dwarfs are known to exhibit periodic radio or $H_{\alpha}$ emission, modulated by the body's rotation period, catalogued in detail in \citet{2013A&A...549A.131A}. The radio emission is believed to be from electron cyclotron maser instability in the brown dwarf magnetosphere \citep{2008ApJ...684..644H,2012ApJ...746...99K}.
Knowing this rotation period and the projected rotational velocity $v\sin{i}$ from spectroscopic observations, it is possible then to determine the radius of the object if its spin is assumed to be perpendicular to the orbital plane of the system.
This has been used by \citet{2009ApJ...695..310B} to determine the radius of a radio- and $H_{\alpha}$- variable component of the brown dwarf binary system 2MASSW J0746425+200032.
This new sample of binary systems therefore also presents the opportunity to systematically study the radii of any brown dwarfs found to have radio emission. Notably, 2M 0036+1821 has a rotation period of $3.08\pm0.05$ hr as determined in \citet{2008ApJ...684..644H}, and therefore presents the clearest new opportunity for radius measurement in this dataset.

\subsection{Binarity Fraction in L Dwarf Sample}
\label{binarity}

We now calculate the revised L-dwarf binary fraction in the 20-parsec sample. Following \citet{2008AJ....135..580R}, correcting for Malmquist bias requires that for statistics in a 20-parsec sample, we should consider only systems within 19 parsecs. The unbiased estimator is then simply the binary fraction observed. For uncertainties, ordinarily Poisson statistics would be used; in this case, however, the sample is too small ($N < 100$), and so binomial statistics are required \citep{2003ApJ...586..512B}.

Where \citet{2008AJ....135..580R} had 8 binaries out of 64 systems observed in the nearest 20 parsecs, giving $\epsilon_b = 12.5^{+5.3}_{-3.0} \%$, five of the previously unresolved systems in the sample now support binary detections. Of these, the spectral class of 2M~2351-2537 is reclassified to L0/L1, while the increased spectroscopic distance to 2M~0045+1634 leaves it outside of 20 parsecs. Given that 2M~2028+0052 was already known to lie beyond 20 pc, 3 of these are admissible in the 20-parsec sample. Furthermore, 2M~1936-5502 supports a high-contrast fit, indicating a potential sub-L-dwarf classification, so among the new discoveries, there are 2 firm detections: 2M~0036+1821 and 2M~2351-2537. In addition to these, we admit a detection of a companion to 2M~0355+1133 in \citet{2010ApJ...715..724B}, for a total of three new binaries. This yields a new binary fraction of $\epsilon_b = 17.2^{+5.7}_{-3.7} \%$. If 2M~1936-5502 is included, the binary fraction is then $18.75^{+5.8}_{-3.7} \%$.

\subsection{Formation of Brown Dwarfs}
\label{formation}
A typical brown dwarf has a mass significantly below the Jeans mass of a collapsing protostellar gas cloud 
\citep{2005nlds.book.....R}, and it is therefore difficult to explain the observed abundance of field brown 
dwarfs based on the standard Jeans collapse theory of star formation. 
This leads to two widely-discussed hypotheses \citep{Basu06072012}. 
The first, namely gravoturbulent collapse, posits that turbulent gas dynamics allows for the collapse of smaller 
clouds than naive analysis suggests, and therefore brown dwarfs may form in essentially the same manner as other stars. 
The alternative is embryo ejection, whereby gravitational interactions in a protostellar system may eject 
low-mass companion embryos before they accrete enough gas to achieve fusion in their cores \citep{MNR:MNR5539}. 
\citet{2005MNRAS.356.1201B} propose that protostellar cores start out at the opacity limit for turbulent 
fragmentation (of order a few $M_J$) and proceed to accrete gas until they are ejected dynamically from the cloud. 
This behavior differs particularly strongly from direct turbulent collapse in the substellar regime, so that 
brown dwarfs are thrown out very early in the accretion process. 
Therefore by testing model predictions for these ultracool stars we provide one of the most stringent tests of
the general validity of the \citet{2005MNRAS.356.1201B} and \citet{2012MNRAS.419.3115B} models of star formation for all stellar types.

The chief objection to direct gravitational collapse models had previously been that the collapse of such a 
low-mass cloud would require it to start out very dense and very cold compared to well-studied star forming 
regions \citep{2005nlds.book.....R}. 
Recent millimeter interferometry by \citet{Andre06072012} has caught just such a gravoturbulent collapse in the 
act, observing a gravitationally bound cloud of mass $\sim 0.02 - 0.03 M_{\astrosun}$. 
In addition this, \citet{2013arXiv1301.4387M}
report the detection of a molecular outflow from the brown dwarf binary FU Tau, which is the third young brown dwarf system found
undergoing a formation process analogous to low-mass stars. 
With at least one example of a brown dwarf forming directly, models that do not account for this behavior 
must therefore be refined. Indeed, simulations by \citet{2012APS..APRC15003J} reproduce the distribution of wide brown 
dwarf companions to main sequence stars as well as the population of close binaries. 

On the other hand, while embryo ejection can easily account for the formation of isolated field brown 
dwarfs, binaries are problematic. 
The binarity fraction of very low mass stars can discriminate between these two models \citep{2007prpl.conf..427B}. 
Embryo ejection predicts a low binarity fraction in field brown dwarfs, as the gravitational interaction required 
would disrupt all but the most tightly bound binaries. 
The predicted binarity fraction is thus $ < 5\%$ in the oldest models for embryo ejection \citep{MNR:MNR5539}, rising only to $8\pm 5\%$ for systems with a primary mass in the range of 0.08 to 0.1 $M_{\astrosun}$ in the most recent models incorporating radiative feedback \citep{2012MNRAS.419.3115B}. 
The observed binarity fraction of ultracool stars seems closer to 
$15\%$ \citep{1538-3881-126-3-1526,2006AJ....132..891R,2008AJ....135..580R} which puts it into conflict with the models of \citet{MNR:MNR5539,2005MNRAS.356.1201B,2012MNRAS.419.3115B}, although the authors noted that this prediction is subject to significant uncertainty. 

The high binarity fraction reported in this paper is still further evidence in favour of a higher binarity rate than predicted by embryo ejection.
Moreover, with the additional detections this prediction now lies several $\sigma$ away from the observed value. 
The hybrid ejection model of \citet{2012ApJ...750...30B}, by which still-collapsing clouds can be ejected from a protostellar disk, makes few clear predictions regarding binarity, other than that there should be few large-separation binaries. Given the results of \citet{Andre06072012} in observing a gravitationally-collapsing brown dwarf mass object, it is not implausible to suggest that direct gravitational collapse may account for most or all field brown dwarfs.

The effect of radiative feedback in ameliorating the difficulties of modelling brown dwarf formation in \citet{2012MNRAS.419.3115B} has been suggested by \citet{2012AJ....144...64D} as an explanation for why the initial mass function and companion mass function changes at the hydrogen burning limit; while this idea is highly speculative, it seems likely that the identification of systems with a late main sequence primary and a brown dwarf secondary as reported in Section~\ref{highcon} will be of interest in testing this hypothesis.

\section{Conclusions}
\label{conclusions}

We have shown that extraction and fitting to the self-calibrating kernel phase interferometric observables allows for 
a significant increase in the robust detection of companions and in the accuracy of best-fit parameters recovered
when compared to simple inspection of images. 
The \emph{Hubble Space Telescope}, or any instrument delivering good wavefront quality may benefit dramatically from
such an approach.

Using the kernel phase technique on all 79 ultracool dwarfs in the combined HST samples of \citet{2006AJ....132..891R} 
and \citet{2008AJ....135..580R}, we independently recover all 10 prior detections and improve on the precision of
fitted parameters such as position angle and separation by a factor of $\sim$10.
Furthermore, we report 5 new binary detections missed by the original authors, 4 of which are presented here for the 
first time, and additionally four marginal detections of close or high contrast companions.
This population forms an excellent base for dynamical studies to establish masses; a prospect particularly
favored by the improved precision in parameter estimation. As well as these confident new detections, kernel phase identifies up to four more marginally-resolved close or faint companions, which may be of planetary mass.

The finding of a larger binary fraction for this sample helps shed light on the formation mechanisms of very low mass objects. 
The five additional ultracool binaries, if confirmed, lend further support to gravoturbulent collapse models 
for the formation of low-mass stars in the field.

\acknowledgements
We would like to thank I. Neill Reid, Olivier Guyon, Paul Stewart, Barnaby Norris, Anthony Cheetham, and James Allison for their interest, helpful feedback and consistent encouragement, as well as Sanjib Sharma for his assistance. We would also like to thank the anonymous reviewer, for drawing attention to several items of recent work and whose comments helped us significantly improve this paper. Support for Program number HST-AR-12849.01-A was provided by NASA through a grant from the Space Telescope Science Institute, which is operated by the Association of Universities for Research in Astronomy, Incorporated, under NASA contract NAS5-26555. This research has made use of NASA's Astrophysics Data System.


\bibliographystyle{aa}

\begin{thebibliography}{61}
\expandafter\ifx\csname natexlab\endcsname\relax\def\natexlab#1{#1}\fi

\bibitem[{{Allen} {et~al.}(2007){Allen}, {Koerner}, {McElwain}, {Cruz}, \&
  {Reid}}]{2007AJ....133..971A}
{Allen}, P.~R., {Koerner}, D.~W., {McElwain}, M.~W., {Cruz}, K.~L., \& {Reid},
  I.~N. 2007, \aj, 133, 971

\bibitem[{{Andr\'{e}} {et~al.}(2012){Andr\'{e}}, {Ward-Thompson}, \&
  {Greaves}}]{Andre06072012}
{Andr\'{e}}, P., {Ward-Thompson}, D., \& {Greaves}, J. 2012, Science, 337, 69

\bibitem[{{Andrei} {et~al.}(2011){Andrei}, {Smart}, {Penna}, {d'Avila},
  {Bucciarelli}, {Camargo}, {Crosta}, {Dapr{\`a}}, {Goldman}, {Jones},
  {Lattanzi}, {Nicastro}, {Pinfield}, {da Silva Neto}, \&
  {Teixeira}}]{2011AJ....141...54A}
{Andrei}, A.~H., {Smart}, R.~L., {Penna}, J.~L., {et~al.} 2011, \aj, 141, 54

\bibitem[{{Antonova} {et~al.}(2013){Antonova}, {Hallinan}, {Doyle}, {Yu},
  {Kuznetsov}, {Metodieva}, {Golden}, \& {Cruz}}]{2013A&A...549A.131A}
{Antonova}, A., {Hallinan}, G., {Doyle}, J.~G., {et~al.} 2013, \aap, 549, A131

\bibitem[{{Aylott} {et~al.}(2009){Aylott}, {Veitch}, \&
  {Vecchio}}]{2009CQGra..26k4011A}
{Aylott}, B., {Veitch}, J., \& {Vecchio}, A. 2009, Classical and Quantum
  Gravity, 26, 114011

\bibitem[{{Baldwin} {et~al.}(1986){Baldwin}, {Haniff}, {Mackay}, \&
  {Warner}}]{1986Natur.320..595B}
{Baldwin}, J.~E., {Haniff}, C.~A., {Mackay}, C.~D., \& {Warner}, P.~J. 1986,
  \nat, 320, 595

\bibitem[{Basu(2012)}]{Basu06072012}
Basu, S. 2012, Science, 337, 43

\bibitem[{{Basu} \& {Vorobyov}(2012)}]{2012ApJ...750...30B}
{Basu}, S. \& {Vorobyov}, E.~I. 2012, \apj, 750, 30

\bibitem[{{Bate}(2012)}]{2012MNRAS.419.3115B}
{Bate}, M.~R. 2012, \mnras, 419, 3115

\bibitem[{{Bate} \& {Bonnell}(2005)}]{2005MNRAS.356.1201B}
{Bate}, M.~R. \& {Bonnell}, I.~A. 2005, \mnras, 356, 1201

\bibitem[{Bate {et~al.}(2002)Bate, Bonnell, \& Bromm}]{MNR:MNR5539}
Bate, M.~R., Bonnell, I.~A., \& Bromm, V. 2002, Monthly Notices of the Royal
  Astronomical Society, 332, L65

\bibitem[{{Berger} {et~al.}(2009){Berger}, {Rutledge}, {Phan-Bao}, {Basri},
  {Giampapa}, {Gizis}, {Liebert}, {Mart{\'{\i}}n}, \&
  {Fleming}}]{2009ApJ...695..310B}
{Berger}, E., {Rutledge}, R.~E., {Phan-Bao}, N., {et~al.} 2009, \apj, 695, 310

\bibitem[{{Bernat} {et~al.}(2010){Bernat}, {Bouchez}, {Ireland}, {Tuthill},
  {Martinache}, {Angione}, {Burruss}, {Cromer}, {Dekany}, {Guiwits}, {Henning},
  {Hickey}, {Kibblewhite}, {McKenna}, {Moore}, {Petrie}, {Roberts}, {Shelton},
  {Thicksten}, {Trinh}, {Tripathi}, {Troy}, {Truong}, {Velur}, \&
  {Lloyd}}]{2010ApJ...715..724B}
{Bernat}, D., {Bouchez}, A.~H., {Ireland}, M., {et~al.} 2010, \apj, 715, 724

\bibitem[{{Bouy} {et~al.}(2003){Bouy}, {Brandner}, {Mart\'{i}n}, {Delfosse},
  {Allard}, \& {Basri}}]{1538-3881-126-3-1526}
{Bouy}, H., {Brandner}, W., {Mart\'{i}n}, E.~L., {et~al.} 2003, The
  Astronomical Journal, 126, 1526

\bibitem[{{Burgasser} {et~al.}(2003){Burgasser}, {Kirkpatrick}, {Reid},
  {Brown}, {Miskey}, \& {Gizis}}]{2003ApJ...586..512B}
{Burgasser}, A.~J., {Kirkpatrick}, J.~D., {Reid}, I.~N., {et~al.} 2003, \apj,
  586, 512

\bibitem[{{Burgasser} {et~al.}(2007){Burgasser}, {Reid}, {Siegler}, {Close},
  {Allen}, {Lowrance}, \& {Gizis}}]{2007prpl.conf..427B}
{Burgasser}, A.~J., {Reid}, I.~N., {Siegler}, N., {et~al.} 2007, Protostars and
  Planets V, 427

\bibitem[{{Chabrier} {et~al.}(2000){Chabrier}, {Baraffe}, {Allard}, \&
  {Hauschildt}}]{2000ApJ...542..464C}
{Chabrier}, G., {Baraffe}, I., {Allard}, F., \& {Hauschildt}, P. 2000, \apj,
  542, 464

\bibitem[{{Dahn} {et~al.}(2002){Dahn}, {Harris}, {Vrba}, {Guetter}, {Canzian},
  {Henden}, {Levine}, {Luginbuhl}, {Monet}, {Monet}, {Pier}, {Stone}, {Walker},
  {Burgasser}, {Gizis}, {Kirkpatrick}, {Liebert}, \&
  {Reid}}]{2002AJ....124.1170D}
{Dahn}, C.~C., {Harris}, H.~C., {Vrba}, F.~J., {et~al.} 2002, \aj, 124, 1170

\bibitem[{{Deleuil} {et~al.}(2008){Deleuil}, {Deeg}, {Alonso}, {Bouchy},
  {Rouan}, {Auvergne}, {Baglin}, {Aigrain}, {Almenara}, {Barbieri}, {Barge},
  {Bruntt}, {Bord{\'e}}, {Collier Cameron}, {Csizmadia}, {de La Reza},
  {Dvorak}, {Erikson}, {Fridlund}, {Gandolfi}, {Gillon}, {Guenther}, {Guillot},
  {Hatzes}, {H{\'e}brard}, {Jorda}, {Lammer}, {L{\'e}ger}, {Llebaria},
  {Loeillet}, {Mayor}, {Mazeh}, {Moutou}, {Ollivier}, {P{\"a}tzold}, {Pont},
  {Queloz}, {Rauer}, {Schneider}, {Shporer}, {Wuchterl}, \&
  {Zucker}}]{2008A&A...491..889D}
{Deleuil}, M., {Deeg}, H.~J., {Alonso}, R., {et~al.} 2008, \aap, 491, 889

\bibitem[{{Dieterich} {et~al.}(2012){Dieterich}, {Henry}, {Golimowski},
  {Krist}, \& {Tanner}}]{2012AJ....144...64D}
{Dieterich}, S.~B., {Henry}, T.~J., {Golimowski}, D.~A., {Krist}, J.~E., \&
  {Tanner}, A.~M. 2012, \aj, 144, 64

\bibitem[{{Evans} {et~al.}(2012){Evans}, {Ireland}, {Kraus}, {Martinache},
  {Stewart}, {Tuthill}, {Lacour}, {Carpenter}, \&
  {Hillenbrand}}]{2012ApJ...744..120E}
{Evans}, T.~M., {Ireland}, M.~J., {Kraus}, A.~L., {et~al.} 2012, \apj, 744, 120

\bibitem[{{Faherty} {et~al.}(2012){Faherty}, {Burgasser}, {Walter}, {Van der
  Bliek}, {Shara}, {Cruz}, {West}, {Vrba}, \&
  {Anglada-Escud{\'e}}}]{2012ApJ...752...56F}
{Faherty}, J.~K., {Burgasser}, A.~J., {Walter}, F.~M., {et~al.} 2012, \apj,
  752, 56

\bibitem[{{Feroz} {et~al.}(2009){Feroz}, {Gair}, {Hobson}, \&
  {Porter}}]{2009CQGra..26u5003F}
{Feroz}, F., {Gair}, J.~R., {Hobson}, M.~P., \& {Porter}, E.~K. 2009, Classical
  and Quantum Gravity, 26, 215003

\bibitem[{{Golimowski} {et~al.}(2004){Golimowski}, {Leggett}, {Marley}, {Fan},
  {Geballe}, {Knapp}, {Vrba}, {Henden}, {Luginbuhl}, {Guetter}, {Munn},
  {Canzian}, {Zheng}, {Tsvetanov}, {Chiu}, {Glazebrook}, {Hoversten},
  {Schneider}, \& {Brinkmann}}]{2004AJ....127.3516G}
{Golimowski}, D.~A., {Leggett}, S.~K., {Marley}, M.~S., {et~al.} 2004, \aj,
  127, 3516

\bibitem[{{Grether} \& {Lineweaver}(2006)}]{2006ApJ...640.1051G}
{Grether}, D. \& {Lineweaver}, C.~H. 2006, \apj, 640, 1051

\bibitem[{{Hallinan} {et~al.}(2008){Hallinan}, {Antonova}, {Doyle}, {Bourke},
  {Lane}, \& {Golden}}]{2008ApJ...684..644H}
{Hallinan}, G., {Antonova}, A., {Doyle}, J.~G., {et~al.} 2008, \apj, 684, 644

\bibitem[{{Hu{\'e}lamo} {et~al.}(2011){Hu{\'e}lamo}, {Lacour}, {Tuthill},
  {Ireland}, {Kraus}, \& {Chauvin}}]{2011A&A...528L...7H}
{Hu{\'e}lamo}, N., {Lacour}, S., {Tuthill}, P., {et~al.} 2011, \aap, 528, L7

\bibitem[{{Ireland}(2013)}]{2013arXiv1301.6205I}
{Ireland}, M.~J. 2013, ArXiv e-prints

\bibitem[{{Ireland} {et~al.}(2008){Ireland}, {Kraus}, {Martinache}, {Lloyd}, \&
  {Tuthill}}]{2008ApJ...678..463I}
{Ireland}, M.~J., {Kraus}, A., {Martinache}, F., {Lloyd}, J.~P., \& {Tuthill},
  P.~G. 2008, \apj, 678, 463

\bibitem[{{Jennison}(1958)}]{1958MNRAS.118..276J}
{Jennison}, R.~C. 1958, \mnras, 118, 276

\bibitem[{{Jumper} \& {Fisher}(2012)}]{2012APS..APRC15003J}
{Jumper}, P. \& {Fisher}, R. 2012, in APS Meeting Abstracts, 15003

\bibitem[{{Kirkpatrick} {et~al.}(2012){Kirkpatrick}, {Gelino}, {Cushing},
  {Mace}, {Griffith}, {Skrutskie}, {Marsh}, {Wright}, {Eisenhardt}, {McLean},
  {Mainzer}, {Burgasser}, {Tinney}, {Parker}, \&
  {Salter}}]{2012ApJ...753..156K}
{Kirkpatrick}, J.~D., {Gelino}, C.~R., {Cushing}, M.~C., {et~al.} 2012, \apj,
  753, 156

\bibitem[{{Kraus} \& {Ireland}(2012)}]{2012ApJ...745....5K}
{Kraus}, A.~L. \& {Ireland}, M.~J. 2012, \apj, 745, 5

\bibitem[{{Kraus} {et~al.}(2011){Kraus}, {Ireland}, {Martinache}, \&
  {Hillenbrand}}]{2011ApJ...731....8K}
{Kraus}, A.~L., {Ireland}, M.~J., {Martinache}, F., \& {Hillenbrand}, L.~A.
  2011, \apj, 731, 8

\bibitem[{{Kraus} {et~al.}(2008){Kraus}, {Ireland}, {Martinache}, \&
  {Lloyd}}]{2008ApJ...679..762K}
{Kraus}, A.~L., {Ireland}, M.~J., {Martinache}, F., \& {Lloyd}, J.~P. 2008,
  \apj, 679, 762

\bibitem[{{Krist} {et~al.}(1998){Krist}, {Golimowski}, {Schroeder}, \&
  {Henry}}]{1998PASP..110.1046K}
{Krist}, J.~E., {Golimowski}, D.~A., {Schroeder}, D.~J., \& {Henry}, T.~J.
  1998, \pasp, 110, 1046

\bibitem[{{Krist} {et~al.}(2011){Krist}, {Hook}, \&
  {Stoehr}}]{2011SPIE.8127E..16K}
{Krist}, J.~E., {Hook}, R.~N., \& {Stoehr}, F. 2011, in Society of
  Photo-Optical Instrumentation Engineers (SPIE) Conference Series, Vol. 8127,
  Society of Photo-Optical Instrumentation Engineers (SPIE) Conference Series

\bibitem[{{Kuznetsov} {et~al.}(2012){Kuznetsov}, {Doyle}, {Yu}, {Hallinan},
  {Antonova}, \& {Golden}}]{2012ApJ...746...99K}
{Kuznetsov}, A.~A., {Doyle}, J.~G., {Yu}, S., {et~al.} 2012, \apj, 746, 99

\bibitem[{{Lloyd} {et~al.}(2006){Lloyd}, {Martinache}, {Ireland}, {Monnier},
  {Pravdo}, {Shaklan}, \& {Tuthill}}]{2006ApJ...650L.131L}
{Lloyd}, J.~P., {Martinache}, F., {Ireland}, M.~J., {et~al.} 2006, \apjl, 650,
  L131

\bibitem[{{Marcy} \& {Butler}(2000)}]{2000PASP..112..137M}
{Marcy}, G.~W. \& {Butler}, R.~P. 2000, \pasp, 112, 137

\bibitem[{{Martinache}(2010)}]{2010ApJ...724..464M}
{Martinache}, F. 2010, \apj, 724, 464

\bibitem[{{Martinache}(2011)}]{2011SPIE.8151E..33M}
{Martinache}, F. 2011, in Society of Photo-Optical Instrumentation Engineers
  (SPIE) Conference Series, Vol. 8151, Society of Photo-Optical Instrumentation
  Engineers (SPIE) Conference Series

\bibitem[{{Martinache} {et~al.}(2007){Martinache}, {Lloyd}, {Ireland},
  {Yamada}, \& {Tuthill}}]{2007ApJ...661..496M}
{Martinache}, F., {Lloyd}, J.~P., {Ireland}, M.~J., {Yamada}, R.~S., \&
  {Tuthill}, P.~G. 2007, \apj, 661, 496

\bibitem[{{Martinache} {et~al.}(2009){Martinache}, {Rojas-Ayala}, {Ireland},
  {Lloyd}, \& {Tuthill}}]{2009ApJ...695.1183M}
{Martinache}, F., {Rojas-Ayala}, B., {Ireland}, M.~J., {Lloyd}, J.~P., \&
  {Tuthill}, P.~G. 2009, \apj, 695, 1183

\bibitem[{{Monin} {et~al.}(2013){Monin}, {Whelan}, {Lefloch}, {Dougados}, \&
  {Alves de Oliveira}}]{2013arXiv1301.4387M}
{Monin}, J.-L., {Whelan}, E., {Lefloch}, B., {Dougados}, C., \& {Alves de
  Oliveira}, C. 2013, ArXiv e-prints

\bibitem[{{Mukherjee} \& {Parkinson}(2008)}]{2008IJMPA..23..787M}
{Mukherjee}, P. \& {Parkinson}, D. 2008, International Journal of Modern
  Physics A, 23, 787

\bibitem[{{Mukherjee} {et~al.}(2006){Mukherjee}, {Parkinson}, \&
  {Liddle}}]{2006ApJ...638L..51M}
{Mukherjee}, P., {Parkinson}, D., \& {Liddle}, A.~R. 2006, \apjl, 638, L51

\bibitem[{{Pecaut} \& {Mamajek}(in preparation)}]{mamajek}
{Pecaut}, M. \& {Mamajek}, E. in preparation, {A Modern Mean Stellar Color and
  Effective Temperatures (Teff) Sequence for O9V-Y0V Dwarf Stars}

\bibitem[{{Pravdo} {et~al.}(2004){Pravdo}, {Shaklan}, {Henry}, \&
  {Benedict}}]{2004ApJ...617.1323P}
{Pravdo}, S.~H., {Shaklan}, S.~B., {Henry}, T., \& {Benedict}, G.~F. 2004,
  \apj, 617, 1323

\bibitem[{{Reid} {et~al.}(2008){Reid}, {Cruz}, {Burgasser}, \&
  {Liu}}]{2008AJ....135..580R}
{Reid}, I.~N., {Cruz}, K.~L., {Burgasser}, A.~J., \& {Liu}, M.~C. 2008, \aj,
  135, 580

\bibitem[{{Reid} \& {Hawley}(2005)}]{2005nlds.book.....R}
{Reid}, I.~N. \& {Hawley}, S.~L. 2005, {New light on dark stars : red dwarfs,
  low-mass stars, brown dwarfs} (New Light on Dark Stars Red Dwarfs, Low-Mass
  Stars, Brown Stars, by I.N.~Reid and S.L.~Hawley.~Springer-Praxis books in
  astrophysics and astronomy.~Praxis Publishing Ltd, 2005.~ ISBN 3-540-25124-3)

\bibitem[{{Reid} {et~al.}(2006{\natexlab{a}}){Reid}, {Lewitus}, {Allen},
  {Cruz}, \& {Burgasser}}]{2006AJ....132..891R}
{Reid}, I.~N., {Lewitus}, E., {Allen}, P.~R., {Cruz}, K.~L., \& {Burgasser},
  A.~J. 2006{\natexlab{a}}, \aj, 132, 891

\bibitem[{{Reid} {et~al.}(2006{\natexlab{b}}){Reid}, {Lewitus}, {Burgasser}, \&
  {Cruz}}]{2006ApJ...639.1114R}
{Reid}, I.~N., {Lewitus}, E., {Burgasser}, A.~J., \& {Cruz}, K.~L.
  2006{\natexlab{b}}, \apj, 639, 1114

\bibitem[{{Sharma} {et~al.}(2011){Sharma}, {Bland-Hawthorn}, {Johnston}, \&
  {Binney}}]{2011ApJ...730....3S}
{Sharma}, S., {Bland-Hawthorn}, J., {Johnston}, K.~V., \& {Binney}, J. 2011,
  \apj, 730, 3

\bibitem[{{Sivaramakrishnan} {et~al.}(2009){Sivaramakrishnan}, {Tuthill},
  {Ireland}, {Lloyd}, {Martinache}, {Soummer}, {Makidon}, {Doyon}, {Beaulieu},
  \& {Beichman}}]{2009SPIE.7440E..30S}
{Sivaramakrishnan}, A., {Tuthill}, P.~G., {Ireland}, M.~J., {et~al.} 2009, in
  Presented at the Society of Photo-Optical Instrumentation Engineers (SPIE)
  Conference, Vol. 7440, Society of Photo-Optical Instrumentation Engineers
  (SPIE) Conference Series

\bibitem[{Sivia \& Skilling(2006)}]{sivia2006data}
Sivia, D. \& Skilling, J. 2006, Data Analysis: A Bayesian Tutorial, Oxford
  Science Publications (OUP Oxford)

\bibitem[{{Skilling}(2004)}]{2004AIPC..735..395S}
{Skilling}, J. 2004, in American Institute of Physics Conference Series, Vol.
  735, American Institute of Physics Conference Series, ed. R.~{Fischer},
  R.~{Preuss}, \& U.~V. {Toussaint}, 395--405

\bibitem[{{Stumpf} {et~al.}(2010){Stumpf}, {Brandner}, {Joergens}, {Henning},
  {Bouy}, {K{\"o}hler}, \& {Kasper}}]{2010ApJ...724....1S}
{Stumpf}, M.~B., {Brandner}, W., {Joergens}, V., {et~al.} 2010, \apj, 724, 1

\bibitem[{{Todorov} {et~al.}(2010){Todorov}, {Luhman}, \&
  {McLeod}}]{2010ApJ...714L..84T}
{Todorov}, K., {Luhman}, K.~L., \& {McLeod}, K.~K. 2010, \apjl, 714, L84

\bibitem[{{Tuthill} {et~al.}(2006){Tuthill}, {Lloyd}, {Ireland}, {Martinache},
  {Monnier}, {Woodruff}, {ten Brummelaar}, {Turner}, \&
  {Townes}}]{2006SPIE.6272E.103T}
{Tuthill}, P., {Lloyd}, J., {Ireland}, M., {et~al.} 2006, in Society of
  Photo-Optical Instrumentation Engineers (SPIE) Conference Series, Vol. 6272,
  Society of Photo-Optical Instrumentation Engineers (SPIE) Conference Series

\bibitem[{{Tuthill} {et~al.}(2000){Tuthill}, {Monnier}, {Danchi}, {Wishnow}, \&
  {Haniff}}]{2000PASP..112..555T}
{Tuthill}, P.~G., {Monnier}, J.~D., {Danchi}, W.~C., {Wishnow}, E.~H., \&
  {Haniff}, C.~A. 2000, \pasp, 112, 555

\end{thebibliography}

\pagebreak

\section{Image and Correlation Diagrams}

The diagrams referred to in the above text are displayed below.

\begin{figure*}[h]
\plottwo{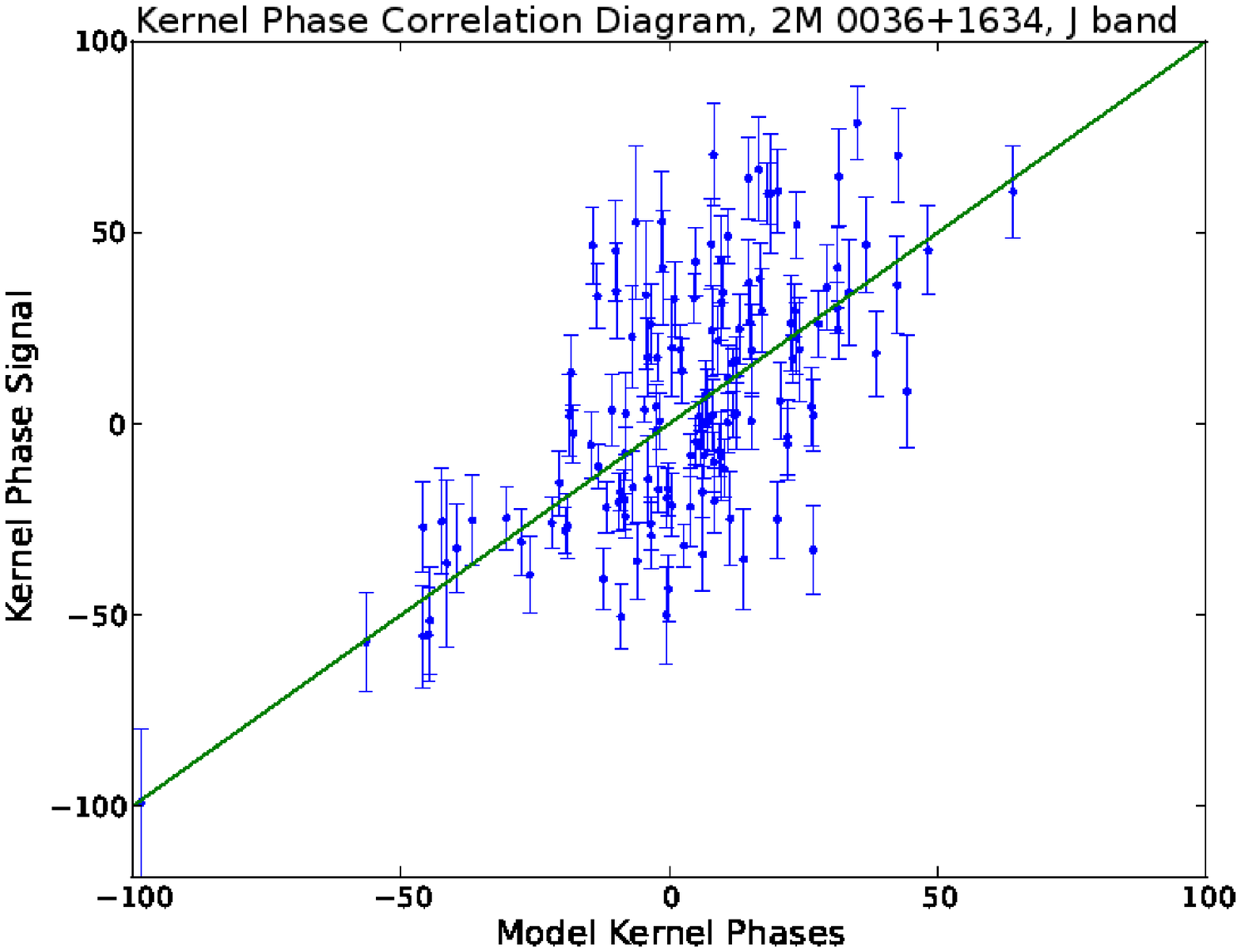}{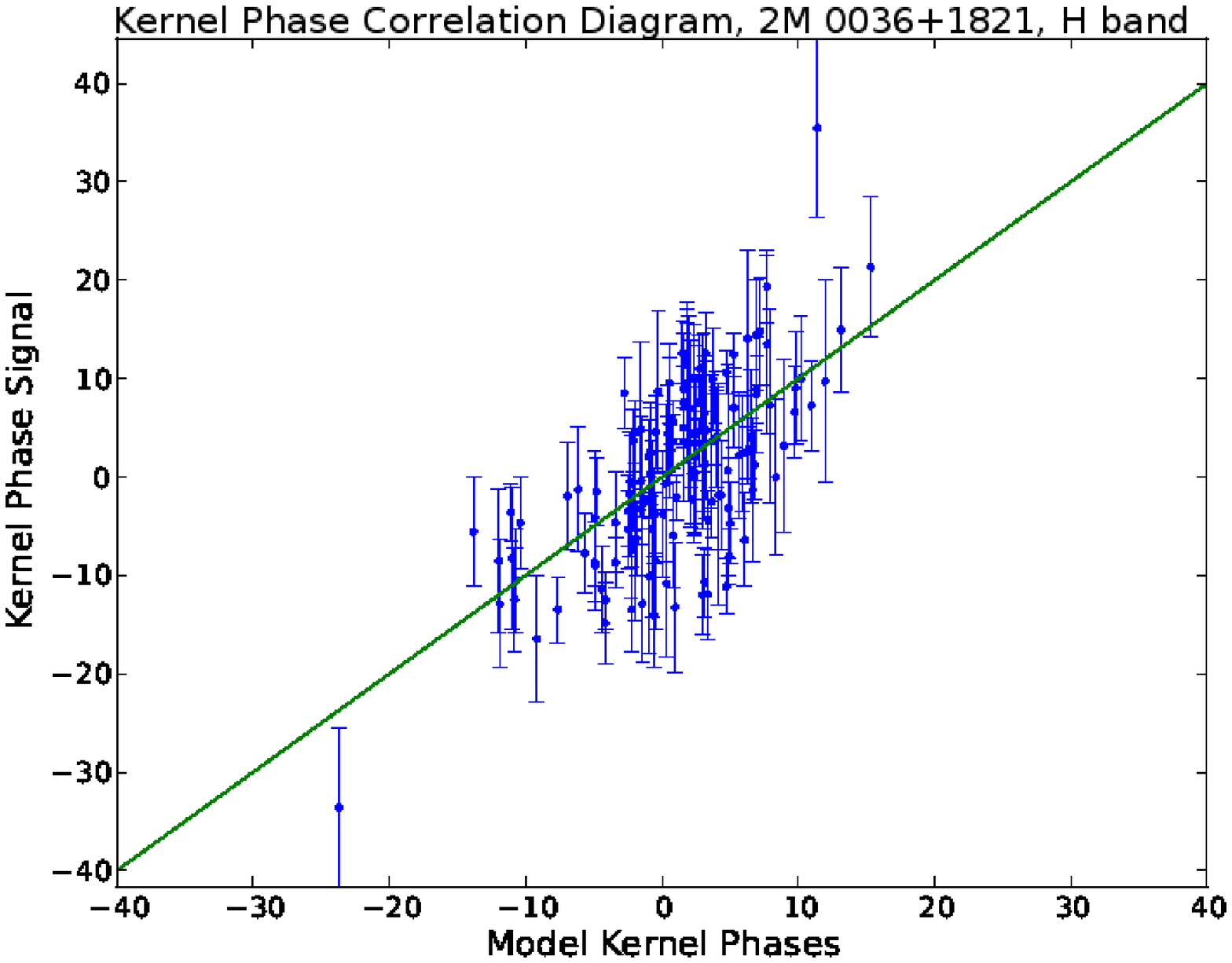}
\caption{Kernel phase correlation diagrams for the newly-confirmed binary 2M 0036+1821 in (left) $J$ band and (right) $H$ band.}
\label{06corr}
\end{figure*}

\end{document}